\documentclass[twocolumn]{aastex62}

\usepackage[utf8]{inputenc}
\usepackage[english]{babel}
\usepackage{hyperref}
\usepackage{amsmath}

\submitjournal{ApJ}

\shorttitle{What Drives Turbulence in Stochastic Reconnection?}
\shortauthors{Kowal et al.}

\begin{document}

\title{Kelvin-Helmholtz versus Tearing Instability: What Drives Turbulence in Stochastic Reconnection?}
\correspondingauthor{Grzegorz Kowal}
\email{grzegorz.kowal@usp.br}

\author[0000-0002-0176-9909]{Grzegorz Kowal}
\affiliation{Escola de Artes, Ci\^encias e Humanidades \\
Universidade de S\~ao Paulo \\
Av. Arlindo B\'ettio, 1000 -- Vila Guaraciaba \\
CEP: 03828-000, São Paulo - SP, Brazil}

\author[0000-0002-1914-6654]{Diego A. Falceta-Gon\c{c}alves}
\affiliation{Escola de Artes, Ci\^encias e Humanidades \\
Universidade de S\~ao Paulo \\
Av. Arlindo B\'ettio, 1000 -- Vila Guaraciaba \\
CEP: 03828-000, São Paulo - SP, Brazil}

\author[0000-0002-7336-6674]{Alex Lazarian}
\affiliation{Department of Astronomy \\
University of Wisconsin,\\
475 North Charter Street, \\
Madison, WI 53706, USA}

\author[0000-0002-2307-3857]{Ethan T. Vishniac}
\affiliation{Department of Physics \& Astronomy \\
Johns Hopkins University \\
3400 N. Charles Street \\
Baltimore, MD 21218, USA}

\begin{abstract}
Over the last few years it became clear that turbulent magnetic reconnection and
magnetized turbulence are inseparable.  It was not only shown that reconnection
is responsible for violating the frozen-in condition in turbulence, but also
that stochastic reconnection in 3D generates turbulence by itself.
The actual mechanism responsible for this driving is still unknown.  Processes
such tearing mode or Kelvin--Helmholtz, among other plasma instabilities, could
generate turbulence from irregular current sheets.
We address the nature of driving mechanism for this process and consider a
relative role of tearing and Kelvin--Helmholtz instabilities for the process of
turbulence generation.  In particular, we analyze the conditions for development
of these two instabilities within three-dimensional reconnection regions.
We show that both instabilities can excite turbulence fluctuations in
reconnection regions.  However, tearing mode has relatively slow growth rate,
and at later times it becomes partially suppressed by transverse to the current
sheet component of magnetic field, generated during the growth of turbulent
fluctuations.  On the contrary, the Kelvin--Helmholtz instability establishes
quickly in the outflow region, and at later times it dominates the turbulence
generation comparing to the contribution from tearing mode.
Our results demonstrate that the tearing instability is subdominant compared to
the the Kelvin--Helmholtz instability in terms of generation of turbulence in
the 3D reconnection layers and therefore the self-driven reconnection is
turbulent reconnection with tearing instability being important only at the
initial stage of the reconnection.
\end{abstract}


\section{Introduction}
\label{sec:intro}

Magnetic reconnection is a fundamental problem essential for understanding the
magnetohydrodynamic (MHD) flows, especially turbulent ones.  Within such flows
magnetic flux tubes cross each other and therefore the properties of the flow
depend on whether the tubes can or cannot cross each other.

The answer that follows from the Sweet--Parker theory of magnetic reconnection
\citep{Parker:1957, Sweet:1958} is that in typical astrophysical situations the
magnetic flux tubes cannot reconnect and change the magnetic field topology.
Indeed, the Sweet-Parker reconnection rate is  $V_\mathrm{rec, SP} \approx
V_\mathrm{A} S_L^{-1/2} \ll V_\mathrm{A}$ with $S_L = L V_\mathrm{A} / \eta$
being the Lundquist number, $L$ - a longitudinal scale of the reconnecting flux
tube, $V_A$ - the Alfv\'en speed, and $\eta$ - the magnetic resistivity.  Given
the large scales of magnetic fields involved in astrophysical flows and the
highly conductive nature of astrophysical plasmas, it is obvious that $S_L$ is
so large that the rates predicted by the Sweet--Parker mechanism are absolutely
negligible.  This, however, is in gross contradiction with observational data,
e.g., the data on Solar flares.  The Sweet--Parker reconnection is an example of
a slow laminar reconnection, while one requires much faster reconnection to
explain observations.  Formally, the fast reconnection is one that does not
depend on $S_L$ or, if depends, depends on it logarithmically.

For years the fast reconnection research was focused on the X-point
reconnection, i.e. the reconnection at which magnetic field is brought at a
sharp angle in the reconnection zone.  This is opposed to the Sweet-Parker
reconnection which is an example of the Y-point reconnection.  The X-point
reconnection was proposed by \cite{Petschek:1964} and required that the inflow
and outflow of the matter into reconnection zone are comparable.  Indeed, the
slow rate of reconnection with Y-point can be viewed as a direct consequence of
the disparity of the scale of astrophysical inflow of the fluid and the scale of
outflow determined by microphysics, i.e., the resistivity or plasma effects, the
latter, however, challenged by \citet[][henceforth LV99]{LazarianVishniac:1999}.

The most significant point of the LV99 theory was that in the presence of the
three-dimensional (3D) turbulence, the reconnection outflow is determined by the
magnetic field wandering and the width of the outflow is the function of
turbulence intensity rather than the resistivity of plasma effects\footnote{The
LV99 proposal was radically different from earlier suggestions of enhancing the
reconnection rate by turbulence.  For instance, \cite{JacobsonMoses:1984}
considered effect of turbulence on micro-scales by increasing Ohmic resistivity.
 Obviously, this could provide the change of the Sweet-Parker rate only by an
insignificant factor.  Similarly, 2D simulations of turbulence in
\cite{MatthaeusLamkin:1985, MatthaeusLamkin:1986} could not shed light on the
actual 3D physics of magnetic reconnection.  Indeed, the component of magnetic
field responsible for the wondering in LV99 model is the Alfv\'enic mode.  This
mode is absent in 2D simulations.  Thus these works were appealing to the
X-points that are produced by turbulence.}.  The predicted by the LV99 theory
dependence of the reconnection rate on the level of turbulence was successfully
tested in the numerical studies of \cite{Kowal_etal:2009, Kowal_etal:2012}.
More recently, these predictions received an additional support from
relativistic MHD simulations by \cite{Takamoto_etal:2015}.  The most important
consequence of the LV99 theory, contrary to all the previous theories of fast
reconnection, was the prediction that the reconnection does not require special
settings, but happens everywhere in turbulent media.  As a result, this violates
flux freezing in astrophysical fluids, which are generically turbulent
\cite[see][]{Eyink:2011, Eyink_etal:2011}.  This remarkable break down of the
classical magnetic flux freezing theorem \citep{Alfven:1942} was numerically
demonstrated in \cite{Eyink_etal:2013}\footnote{The violation of flux freezing
in turbulent fluids entails an important effect of reconnection diffusion that
has big consequences, changing the paradigm of magnetically controlled star
formation \cite[see][]{Lazarian:2005, Santos-Lima_etal:2010,
Lazarian_etal:2012}.}.

Turbulence can be both, externally driven, as it is testified from the
observations of the ISM and molecular clouds \cite[see][]{Armstrong_etal:1995,
Padoan_etal:2009, ChepurnovLazarian:2010, Chepurnov_etal:2015}, but it can also
be driven by the reconnection process as first discussed in LV99 and further
elaborated in \cite{LazarianVishniac:2009}.  The first numerical study of
magnetic reconnection induced by turbulence that is generated by reconnection
were performed in \cite{Beresnyak:2013} with an incompressible code, and later
in \cite{Oishi_etal:2015} and \cite{HuangBhattacharjee:2016} taking into account
compressibility.  A detailed numerical study of reconnection with self-generated
turbulence was performed in \cite{Kowal_etal:2017}.

One of the most important questions of the current research on 3D reconnection
faces the nature of turbulence in the reconnection events.  Our earlier study in
\cite{Kowal_etal:2017} demonstrated that the turbulence generated in the
reconnection events follows the Goldreich--Sridhar statistics
\citep{GoldreichSridhar:1995}.  However, an open issue is related to the driving
mechanism of the observed turbulent motions.  The literature has suggested that
tearing modes, plasmoid instabilities, and shear-induced instabilities could
mediate the energy transfer from coherent to turbulent flows.  The issue of the
relative importance of different driving processes has not been explored
quantitatively.

In our numerical experiments we do not identify tearing modes, although
filamentary plasmoid-like structures are present.  Visual inspection shows,
however, that their filling factor is small.  Sheared flows, on the other hand,
are present around and within the whole current sheet.  As the field lines
reconnect, the Lorentz force increases, accelerating plasma and creating the
current sheet.  This process is patchy and bursty in 3D.  Therefore, the
accelerated flows are strongly sheared.  The statistical importance of these
burst flows is large, as already shown in our previous work
\citep{Kowal_etal:2017}, as we compared the velocity anisotropy of reconnecting
events to that of decaying turbulence without the reversed field.  The
Kelvin--Helmholtz instability due to the sheared velocities in reconnecting
layers has already been conjectured as possible origin of turbulence by
\cite{Beresnyak:2013}.  In \cite{Kowal_etal:2017}, we provided the solid
evidence for the self-generated turbulence driven by the velocity shear.  Here
we perform a proper analysis of the growth-rates of such instabilities.

Velocity shear is a global process that occurs in regular magnetized and
unmagnetized fluids.  The nonlinear evolution of the related instabilities, such
as Kelvin--Helmholtz instability, is known to be one of the main contributors to
the energy transfer between wave modes, i.e., the energy cascade.  If the energy
cascade in reconnection layers is led by similar mechanisms, it is
straightforward to understand why the statistics observed resemble those of
Kolmogorov-like turbulence, and Goldreich--Sridhar anisotropy scaling.  In other
words, our claim is that the turbulent onset and cascade in reconnection is not
different to those found in regular MHD and hydrodynamic systems.

In this work we consider two possible instabilities both extensively studied
analytically and numerically since 1960s, namely tearing mode instability in a
slab geometry \citep[see][]{Furth_etal:1963, SomovVerneta:1989}, which naturally
develops in a thin elongated current sheets, and Kelvin--Helmholtz instability
\citep[see, e.g.][]{Chandrasekhar:1961, Michalke:1964, Fejer:1964, Sen:1964,
Gerwin:1968, OngRoderick:1972, MiuraPritchett:1982, Frank_etal:1996,
FaganelloCalifano:2017, BerlokPfrommer:2019}, which could result from the local
velocity shear produced by the outflows from reconnection sites.  Both
instabilities are able to generate turbulence near current sheets. In this work
we investigate which one is responsible for or dominates the generation of
turbulence in stochastic reconnection, i.e. the reconnection without an
externally imposed turbulence resulting from a weak initial plasma
irregularities.

In what follows we provide a brief review of theoretical analysis of
Kelvin-Helmholtz and tearing instabilities in \S\ref{sec:instabilities},
together with dispersion relations derived for different magnetic field and
velocity profiles and conditions for their suppression.  In \S\ref{sec:model},
we describe numerical simulation and our approach in analyzing both
instabilities.  In \S\ref{sec:results} we present the results of our analyses
and compare the maximum growth rates of the two instabilities at different
times, which we discuss in \S\ref{sec:discussion}.  Finally, in
\S\ref{sec:conclusions} we state our conclusions.

\section{Analyzed Instabilities}
\label{sec:instabilities}

%
\subsection{Tearing Mode Instability}
\label{ssec:tr-analysis}

With tokamaks in mind, many works on tearing mode instability considered a
cylindrical plasma column in so called ``pinch configuration'' or simply
toroidal geometry.  In such a configuration, the current density pinch, due to
the change of the azimuthal component of magnetic field, is located at some
distance from the column center and the axial component is usually stronger than
the azimuthal one \cite[see, e.g.][]{Coppi_etal:1976, Ara_etal:1978}.  In this
subsection we briefly review the dispersion relations for tearing mode derived
in a slab geometry, where one component of magnetic field changes the sign over
a short distance $2 \delta$ along its perpendicular direction.  It is know that
independently of the configuration, the plasma is stable in the ideal MHD
framework.  The stability occurs once a finite resistivity is taken into
account.

\cite{Furth_etal:1963} developed a theory of resistive MHD instabilities for the
case of the neutral current layer with an arbitrary magnetic field profile.
However, in derivation of the dispersion relations they considered a piecewise
linear or hyperbolic tangent profiles.  From their theory, we know that the
tearing mode develops under condition $\alpha = k \delta < 1$, where $k$ is the
perturbation wavenumber (along the sheet plane) and $\delta$ is the current
sheet half-width \cite[see Table 1 in][]{Furth_etal:1963}.  Under the assumption
of the so called ''constant $\Psi$'' in the region of discontinuity \cite[see
\S5 in][]{Furth_etal:1963}, where $\Psi$ is the flux function of the in-plane
magnetic field, $\vec{B} = \left( - \partial_y \Psi, \partial_x \Psi \right)$,
the growth rate $\gamma$, expressed by the renormalized growth rate $p = \gamma
\tau_R$, of the instability for the piecewise linear profile is given by
\begin{equation}
p \approx \alpha^{-2/5} S^{2/5}
\label{eq:tr_grate_furth_short}
\end{equation}
with $S = {\tau_R} / {\tau_A}$, where $\tau_R = {4 \pi \delta^2} / {\eta}$ is
the resistive diffusion time, $\tau_A = {\delta} / {v_A}$ is the Alfv\'en time,
and $v_A = {B_0} / {\sqrt{4 \pi \rho_0}}$ is the upstream Alfv\'en speed.  This
growth rate corresponds to short-wavelength modes, $\alpha S^{1/4} \ll 1$.  For
the long-wavelength modes, $\alpha S^{1/4} \gg 1$, the function $\Psi$ is no
longer constant in the region of discontinuity and the dispersion relation takes
form
\begin{equation}
p \approx \alpha^{2/3} S^{2/3}.
\label{eq:tr_grate_furth_long}
\end{equation}
The existence of these two regimes with different scalings indicates that the
growth rate of tearing mode has its maximum $p_{max}$ with respect to $\alpha$,
which according to \cite{Furth_etal:1963} scales as $p_{max} \sim S^{1/2}$.
\cite{Coppi_etal:1976}, analyzing the internal kink instability of a plasma
column, and \cite{Loureiro_etal:2007}, considering a finite length current
sheet, derived complete dispersion relations covering both regimes.

The tearing mode usually is studied in the two-dimensional (2D) incompressible
MHD framework, and since the theory by \cite{Furth_etal:1963} there was search
for additional mechanism which could stabilize or destabilize ever more the
instability. \cite{Furth_etal:1963} already considered effects of fluid
compressibility, thermal conductivity or finite viscosity concluding that they
do not affect the instability.  Nevertheless, the tearing instability is a
subject to suppression under some circumstances.  For instance, the presence of
traverse, i.e. normal to the current sheet plane, magnetic field component
stabilizes tearing mode.  The stabilization effect of this component was
demonstrated by \cite{Schindler:1974} and \cite{GaleevZelenyi:1975,
GaleevZelenyi:1976}.

\cite{SomovVerneta:1989} have shown that the stabilization effect of the
transverse field $B_n$ becomes to be essential when $\xi \equiv {B_n}/{B_0} \geq
S^{-3/4}$, decreasing the growth rate $p$ with increasing the value of $\xi$.
Once $\xi \gg S^{-3/4}$, the instability is completely stabilized. They derived
the dispersion relation for the tearing instability in the presence of $B_n$,
which is given by
\begin{equation}
\Delta^{1/4} \left( \frac{\alpha^2 S^2}{p} \right)^{1/4}
\left( 1 - \frac{p^{3/2}}{\alpha S} \Delta^{-1/2} \right)
 - p \alpha \sqrt{\frac{\pi}{2}} = 0,
\label{eq:tr_grate_xi_full}
\end{equation}
where $\Delta = \left( 1 + {\xi^2 S^2} / {p} \right)^{-1}$.  The above
dispersion relation is valid for $\xi < \alpha$.  Once $\xi \rightarrow 0$
($\Delta \rightarrow 1$), Eq.~(\ref{eq:tr_grate_xi_full}) reduces to
\begin{equation}
\left( \frac{\alpha^2 S^2}{p} \right)^{1/4}
\left( 1 - \frac{p^{3/2}}{\alpha S} \right)
 - p \alpha \sqrt{\frac{\pi}{2}} = 0,
\end{equation}
properly recovering the scalings in the short- and long-wavelength regimes
discussed before. Applying an implicit derivative to
Eq.~(\ref{eq:tr_grate_xi_full}) with respect to $\alpha$ and setting $dp/d\alpha
= 0$ provides the following relationships between the maximum growth rate
$p_{max}$ and the corresponding $\alpha_{max}$
\begin{multline}
\alpha_{max} = \frac{2}{\sqrt[3]{\pi}} \left( \frac{p_{max}}{S^2}
 + \xi^2 \right)^{1/6} \quad \mathrm{and} \\ \quad p_{max}
 = \left( \frac{\pi^2}{64} \alpha_{max}^6 - \xi^2 \right) S^2.
\label{eq:tr_max}
\end{multline}
Inserting the equation for $\alpha_{max}$ into the dispersion relation
Eq.~(\ref{eq:tr_grate_xi_full}) gives us an expression for the maximum growth
rate $p_{max}$, which depends only on $\xi$ and $S$,
\begin{equation}
p_{max}^4 + \xi^2 S^2 p_{max}^3 - \frac{1}{\pi} \left( \frac{2}{3} \right)^3 S^2 = 0.
\label{eq:tr_max_xi}
\end{equation}
The resulting expression is a quartic function which has negative determinant
for any possible value of $\xi$, thus it has two real roots, one negative and
one positive, with the positive one being the physical maximum growth rate.
Provided the estimates of $\xi$ and $S$ from the simulations, we can therefore
determine the maximum growth rate $p_{max}$.  For $\xi = 0$ this equation gives
directly the maximum growth rate $p_{max} = \left( {2}/{3} \right)^{3/4}
\pi^{-1/4} S^{1/2}$ compatible with the estimate given by
\cite{Furth_etal:1963}.  Once $p_{max}$ is determined, we can also retrieve the
corresponding parameter $\alpha_{max}$ from the left expression in
Eq.~(\ref{eq:tr_max}), from which we can calculate the wavenumber $k_{max} =
\alpha_{max} / \delta$ at which the maximum growth rate $p_{max}$ takes place.

%
\subsection{Kelvin--Helmholtz Instability}
\label{ssec:kh-analysis}

The Kelvin--Helmholtz instability can develop in a region of velocity shear.
Following \citet[][ \S XI, Eq. 204]{Chandrasekhar:1961}, we can write its growth
rate $\gamma$ for an incompressible flow, assuming equal density on both sides
and perturbation parallel to the shear direction, as
\begin{equation}
 \gamma = k \sqrt{U^2 - v_A^2}, \quad U > v_A
 \label{eq:kh_grate_disc}
\end{equation}
where $U$ is the velocity shear amplitude, assuming it changes from $-U$ to $U$
between two layers, $v_A$ is the local Alfvén speed and $k$ is the wavenumber of
perturbation.  This expression, however, is valid only for a simplified case of
a ``vortex sheet'', where the velocity has a discontinuity in the sheet
location, and gives unrealistic growth rate, increasing with $k$ to arbitrarily
large values.  A more realistic continuous shear layer was already considered
long time before.  In the hydrodynamic framework, \cite{Rayleigh:1879} derived a
dispersion relation for a linear piecewise profile of velocity.  Also
\cite{Chandrasekhar:1961} considered continuous variations of density and
velocity profiles (see \S102--104 and Fig.~120 there), while
\cite{Michalke:1964} analyzed numerically an hyperbolic tangent profile.  All
these works concluded that the growth rate reaches its maximum value
$\gamma_{max}$ for $k_{max} \delta < 1$, where $\delta$ is the half-width of the
shear region.  Following \cite{Rayleigh:1879} derivation, \citet[][ see
\S23]{DrazinReid:1981} provided the growth rate for the Kelvin--Helmholtz
instability with a piecewise linear velocity profile in the hydrodynamic
framework,
\begin{equation}
\gamma^2 = \frac{U^2}{4 \delta^2} \left[ e^{-4 \alpha}
 - \left( 1 - 2 \alpha \right)^2 \right],
\label{eq:kh_grate_hd}
\end{equation}
where $\alpha \equiv k \delta$.  The expression indicates that the instability
is completely stabilized for $\alpha \gtrsim 0.639$, and the maximum growth rate
occurs at $\alpha \approx 0.398$.  In the incompressible MHD framework,
\cite{OngRoderick:1972}, considering also a piecewise linear profile, derived
the following dispersion relation
\begin{multline}
\Omega^4 + \Omega^2 \left[ \alpha - 2 \alpha^2 \left( 1 + 1 / {\cal M_A}^2 \right)
 - \frac{1}{4} \left( 1 - e^{-4 \alpha} \right) \right] \\
 + \alpha^4 \left( 1 - 1 / {\cal M_A}^2 \right)^2 - \alpha^3
 \left( 1 - 1 / {\cal M_A}^2 \right) \\ + \frac{1}{4} \alpha^2
 \left( 1 - e^{-4 \alpha} \right) = 0,
\label{eq:kh_grate_imhd}
\end{multline}
where $\Omega = \omega \delta / U$, ${\cal M_A}^2 = U^2 / v_A^2$ is the
Alfv\'enic Mach number related to the shear amplitude $U$, and $\omega = \phi +
i \gamma$, with $\phi$ and $\gamma$ being the phase change and growth rate,
respectively. The dispersion relation expressed by Eq.~(\ref{eq:kh_grate_imhd})
is valid, however, in the limit ${\cal M_A}^2 \gg 1$ for $\alpha \le 1.0$.  In
the hydrodynamic limit, i.e., once ${\cal M_A}$ tends to infinity, the
dispersion relation becomes compatible with the relation provided by
\cite{DrazinReid:1981}.  Similar analytical studies of magnetized
Kelvin-Helmholtz instability with a linear velocity profile were done by
\cite{GudkovTroshichev:1996} resulting in compatible conclusions.

\cite{OngRoderick:1972} also studied the Kelvin-Helmholtz instability in the
compressible MHD framework.  In this regime, they derived the following
dispersion relation
\begin{multline}
8 \alpha \mu^2 \Omega^4 + \Omega^2 \left[ - 4 + \mu^2
 \left( - 1 - 4 \alpha + 8 \alpha^2 - \frac{10}{8} \alpha^3 \right) \right] \\
 + \left( 1 - 2 \alpha \right)^2 - e^{-4 \alpha} \\ - \mu^2
 \left( \alpha^2 - \frac{4}{3} \alpha^3 - \frac{8}{3} \alpha^4
 + \frac{8}{5} \alpha^5 \right) = 0,
\label{eq:kh_grate_mhd}
\end{multline}
valid when $\mu^2 \equiv {\cal M}^2 / \alpha^2 \ll 1$ with ${\cal M} = U / c_M$,
$c_M = \sqrt{a^2 + v_A^2}$, and $a$ being the magnetosonic Mach number, the
magnetosonic speed, and the sound speed, respectively.  Once $\mu^2 \rightarrow
0$, Eq.~(\ref{eq:kh_grate_mhd}) reduces to the relation compatible with
Eq.~(\ref{eq:kh_grate_hd}).  Both above dispersion relations
(Eqs.~\ref{eq:kh_grate_imhd} and \ref{eq:kh_grate_mhd}) indicate the existence
of the maximum growth rate $\gamma_{max}$ at corresponding wavenumber $k_{max}$
also in the magnetized version of the Kelvin-Helmholtz instability, with
$k_{max} \delta \lesssim 1$.  In order to find $\gamma_{max}$ and $k_{max}$, one
has to solve these relations numerically for a given $\delta$, $U$, $v_A$, and
$a$, which is somewhat simplified since both equations are biquartic functions.

\cite{MiuraPritchett:1982} studied the Kelvin-Helmholtz instability numerically
in a general configuration, i.e., taking into account the compressibility and
arbitrary angles between the shear direction, magnetic field and perturbation
propagation.  They derived a sufficient condition for stability, $U \le v_A
\left( \vec{k} \cdot \hat{B}_0 \right) / \left( \vec{k} \cdot \hat{U} \right)$
which, assuming that the perturbation wavevector $\vec{k}$ and magnetic field
$\vec{B}_0$ both lay in the plane of the maximum shear and form angles $\phi$
and $\theta$ with the shear direction, respectively, can be written as
\begin{equation}
U \le v_A \left( \cos \theta + \sin \theta \tan \phi \right).
\label{eq:kh-condition}
\end{equation}
The condition states that if the magnetic field is aligned with the shear
($\vec{B}_0 \parallel \vec{U}$, i.e. $\theta = 0$), the angle of the
perturbation propagation $\phi$ does not matter and the mode is stable if $U \le
v_A$.  Once the magnetic field is perpendicular to the shear ($\vec{B}_0 \perp
\vec{U}$, i.e. $\theta = 90^\circ$), the perturbations are stabilized only for
modes nearly perpendicular to the shear ($\phi \approx 90^\circ$).  Moreover,
considering the perturbation propagating strictly along the shear ($\vec{k}
\parallel \vec{U}$), for the perpendicular field ($\vec{B}_0 \perp \vec{U}$),
the instability develops only if ${\cal M} < 1$, which is equivalent to $U <
c_M$.  For the parallel magnetic field ($\vec{B}_0 \parallel \vec{U}$), the
instability is completely suppressed for ${\cal M}_s = U/a > 1$, once ${\cal
M}_A > 1$ has to be fulfilled for
instability.\footnote{\cite{MiuraPritchett:1982} considered the velocity change
from $-V_0/2$ to $V_0/2$ and calculated Mach numbers using $V_0$ resulting in
respective conditions for Mach numbers ${\cal M}_s < 2$ and ${\cal M}_A > 2$.}
However, the restrictions for ${\cal M}$ and ${\cal M}_s$ are relaxed for
perturbations propagating obliquely with respect to the sheared velocity
component and magnetic field \cite[see Eq. 37 in][]{MiuraPritchett:1982}.
Moreover, it is important to notice that even though the system may be strongly
magnetized overall, in the regions where reconnection occurs the local degree of
magnetization decreases considerably, allowing the growth of KH-unstable modes
\cite[see, e.g.,][]{Loureiro_etal:2013}.

\section{Methodology and Modeling}
\label{sec:model}

%
\subsection{Numerical Simulations}
\label{ssec:simulations}

In this work we analyze a numerical simulation of the reconnection-driven
turbulence in a 3D rectangular domain with physical dimensions $L \times 4L
\times L$ (with assumed $L = 1$) by solving  non-ideal isothermal compressible
magnetohydrodynamic equations using a high-order shock-capturing adaptive
refinement Godunov-type code AMUN\footnote{The code is freely available from
\href{http://amuncode.org}{http://amuncode.org}}.  The exact numerical setup of
the model analyzed here is the same as of the models presented in
\citet{Kowal_etal:2017}, with the sound speed $a = 1.0$ ($\beta = 2.0$), and
explicit viscosity $\nu$ and resistivity $\eta$, both equal to $10^{-5}$, in
order to control the effects of numerical diffusion.  The model was ran with the
base resolution of $32 \times 128 \times 32$, and the local mesh refinement up
to $7$ levels, with the refinement criterion based on the normalized value of
vorticity and current density, resulting in the effective resolution $2048
\times 8192 \times 2048$ or the effective grid size $h = 1 / 2048$ (the same for
all directions).  The initial density $\rho_0$ and the strength of reconnecting
component of magnetic field $B_0$ were set to unity, resulting in the velocity
and the simulation time units being $[ v ] = V_A = 1$ and $[t] = t_A = L / V_A =
1$, respectively.  It is important to note, that the code works with the
normalized magnetic field multiplied by a factor of $\sqrt{4 \pi}$.  Therefore,
the Alfv\'en speed using the code units is given by $V_A = B_0 / \sqrt{\rho_0}$.
In addition to the reconnecting magnetic field component along the X-direction,
we also set the uniform guide field along the Z-direction of the strength of
$0.1$. The velocity field was initiated with fluctuations represented by $100$
Fourier modes of random phases and directions, the amplitude equal to $10^{-2}$
and the fixed wavenumber of $k = 64 \pi$.  The simulation was terminated after
$t = 7.0$.  For a complete description of the numerical setup of the model, the
boundary conditions and numerical methods used please refer to
\citet{Kowal_etal:2017}.

%
\subsection{Shear Detection in Vector Fields}
\label{ssec:shear_detection}

In order to detect locations of the current sheet we analyze a quantity which is
correlated with the local change of the polarization of magnetic field, or
simply the magnetic shear.  There are several techniques proposed in the
literature to determine locations where the reconnection takes place.  The most
straightforward is the amplitude of current density $|\vec{J}|$.  We can also
use the magnetic shear angle, i.e. the change of the magnetic field direction
along a line, or the Partial Variance of Increments method (PVI) which measures
the variation of the magnetic field across the current sheet
\cite[see][]{Greco_etal:2008,Servidio_etal:2011}.  Similarly, for the velocity
shear, we can consider, for example, the vorticity $\vec{\omega} = \nabla \times
\vec{v}$ as the shear detector.  In this work we analyze the local maximums of
the euclidean norm of the shear rate tensor $S_{ij} = {\partial u_i}/{\partial
x_j} + {\partial u_j}/{\partial x_i}$ for $i \ne j$ and $S_{ij}=0$ for $i=j$, $S
= \|S_{ij}\| \equiv \sqrt{\sum_{ij} S_{ij}^2}$, where $x_i, x_j \in \{x, y, z\}$
and $\vec{u}$ means either $\vec{B}$ or $\vec{v}$, depending on the analyzed
instability. Therefore, $S = S(x,y,z)$ is a function of position.

\begin{figure*}[t]
\centering
\includegraphics[width=0.48\textwidth]{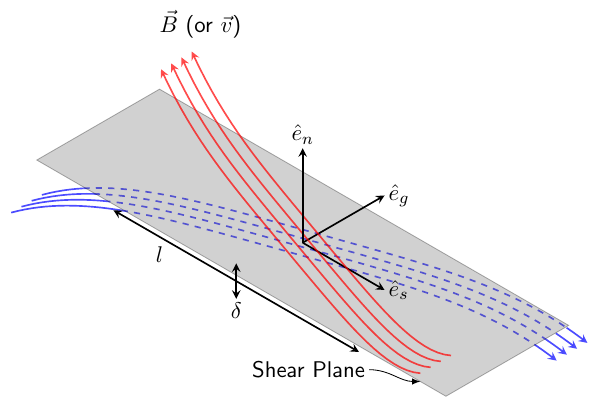}
\includegraphics[width=0.48\textwidth]{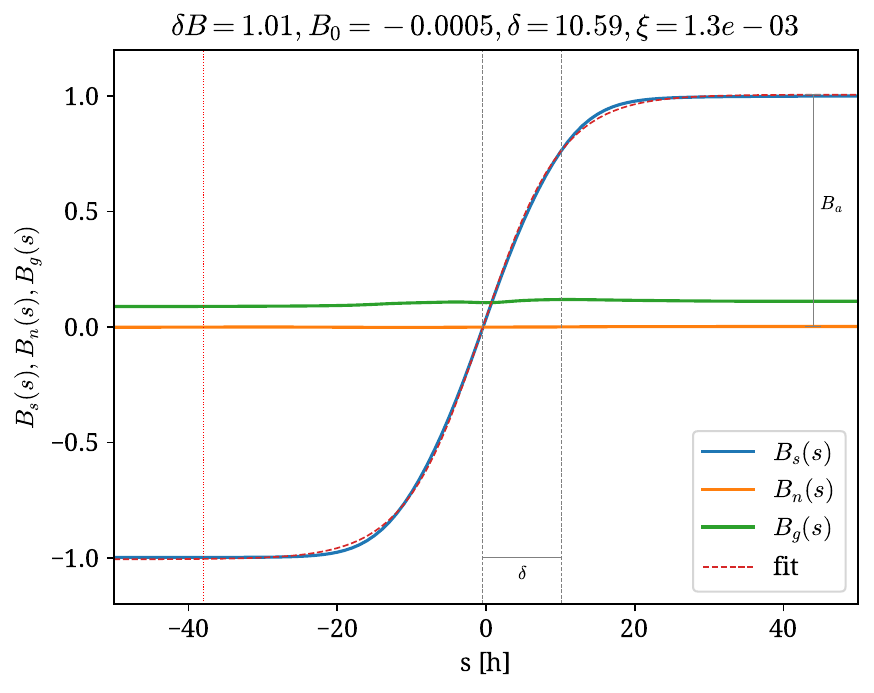}
\caption{{\it Left:} Sketch of a shear region with the local reference frame
indicating directions of the shear $\hat{e}_s$, transverse (normal) $\hat{e}_n$,
and guide components $\hat{e}_g$ of the analyzed vector field ($\vec{B}$ or
$\vec{v}$ field).  The length $l$ of the shear region and its thickness $\delta$
are determined along $\hat{e}_s$ and $\hat{e}_n$ axes, respectively.
{\it Right:} Example profiles of the shear (reconnecting), transverse and guide
components of magnetic field ($B_s$, $B_n$, and $B_g$, respectively) projected
on the direction normal to current sheet in one of the selected points of the
detected current sheet. The estimated parameters of the fitting of the
reconnecting component $B_s(s)$ are shown in the title. The horizontal scale is
in the units of the effective cell size $h$.
\label{fig:fit}}
\end{figure*}

Our algorithm to determine the local geometry of the shear consists of the
following steps:
\begin{enumerate}
\item At each domain cell $\vec{p} = (x,y,z)$ we calculate the 2$^{nd}$ order
partial derivatives of $S(x,y,z)$ along each direction, $\partial_x^2 S$,
$\partial_y^2 S$, and $\partial_z^2 S$ in order to verify if the position is a
local maximum.  The type of maximum is determined by the number of negative
2$^{nd}$ order derivatives. In a given point of the 3D domain, one, two or three
negative 2$^{nd}$ order derivatives identify a sheet-like, ridge-like, or
peak-like local maximums, respectively.  In our analysis, we consider all types
of the local maximums, requiring that at least one derivative is negative.  If
all derivatives are non-negative, the remaining steps of analysis are skipped
and the next cell is considered.  This guarantees that the next computationally
expensive steps are avoided for cells which do not lay in the regions of shear
maximums.

\item For each cell of the local shear maximum, i.e., the cell which has at
least one negative 2$^{nd}$ order derivative determined in the previous step, we
calculate the complete Hessian of the analyzed shear detector $S$
\begin{equation}
 H_{ij}^J = \frac{\partial^2 S}{\partial x_i \partial x_j},
\end{equation}
where again $x_i, x_j \in \{x, y, z\}$.  By solving the eigenproblem for the
Hessian, we can get the direction of the steepest decay of $S$ at the given
location, which is determined by the eigenvector $\hat{e}_{n} =
\hat{e}(\lambda_\mathrm{min})$ corresponding to the minimum eigenvalue
$\lambda_\mathrm{min} = \min{\{\lambda_i\}}$.  The previous step guarantees the
existence of at least one negative eigenvalue.

\item The eigenvector $\hat{e}_{n}$ is perpendicular to the component which
suffers the strongest shear.  However, we would like to determine a complete
vector base corresponding to the directions along the strongest shear,
$\hat{e}_{s}$, across that shear, $\hat{e}_{n}$, and perpendicular to both,
$\hat{e}_{g} = \hat{e}_{n} \times \hat{e}_{s}$ (see the left panel of
Fig.~\ref{fig:fit}).  Since a pure shear is a combination of the normal stress
and rotation, is it not trivial to determine its precise direction.  Moreover,
in a generic case, the stress may not be planar, acting along all three
principal axes.  In order to deal with such a situation, we can assume that we
are only interested in a plane defined by two principal axes, corresponding to
the strongest positive and strongest negative normal stresses. In order to
determine them, we need to solve the eigenproblem for the shear rate tensor
$S_{ij}$.  The eigenvector, $\hat{e}_{g}$, corresponding to the middle
eigenvalue determines the plane of the maximum shear.  Therefore, the direction
of the sheared component $\hat{e}_{s}$ is
\begin{equation}
 \hat{e}_s = \hat{e}_n \times \hat{e}_{g}.
\end{equation}
There is no reason for $\hat{e}_{n}$ and $\hat{e}_{g}$ to be perfectly
orthogonal.  Therefore, after normalizing $\hat{e}_{s}$, $\hat{e}_n$ is
corrected by setting it to $\hat{e}_s \times \hat{e}_{g}$.  This completes the
determination of the orthonormal base of the shear.

\item After determining our orthonormal base in the local shear maximum, we
perform the interpolation of all three components of the analyzed vector field,
$u_x$, $u_y$, and $u_z$ along the normal direction determined by vector
$\hat{e}_{n}$ within a distance of several cell sizes (e.g., $s = {-20h, \dots,
20h}$, where $s$ is the distance in the units of cell size $h$).  Projecting the
resulting vectors on $\hat{e}_{s}$ we obtain the sheared component of the
analyzed vector field,
\begin{equation}
u_s(s) = \hat{e}_{s} \cdot \vec{u}(\vec{p} + s \hat{e}_{n}).
\end{equation}
For the interpolation, we use a piecewise quintic Hermite interpolation which
preserves the continuity of the first and second derivatives \cite[see,
e.g.,][]{Dougherty_etal:1989}.

\item At this point we perform fitting of the sheared component $u_s(s)$ to the
hyperbolic tangent function $f(s) = \delta u \tanh\left[(s - s_0) / \delta
\right] + u_0$ (or a piecewise linear function) to the obtained profile
$u_s(s)$.  The fitting is done within the region $s_l \le s \le s_u$, where $s_l
< 0$ and $s_u > 0$, and $s_l$ and $s_u$ indicate distances along $\hat{e}_{s}$
at which the derivative ${du_s}/{ds}$ changes sign.  The estimated values
$\delta u$,  $u_0$ correspond to the amplitudes of the sheared and uniform
components, respectively, $s_0$ is the shift with respect to the current
position, and $\delta$ is the half-width of the profile.  The right panel in
Figure~\ref{fig:fit} shows an example of the determined sheared component of
magnetic field $B_s(s)$ together with its fitting (blue and dashed red lines,
respectively).  The fitting parameters are shown in the title of the plot.

\item Along the normal direction, $\hat{e}_{n}$, we can also project other
quantities.  For example, in the case of the tearing mode we estimate the
transverse component of magnetic field $B_n(s) = \hat{e}_{n} \cdot
\vec{B}(\vec{p} + s \hat{e}_{n})$.  By averaging $B_n(s)$ within the local
current sheet, i.e. within the interval $s_l \le s \le s_u$ of the fitted
function, we can get the mean value of the transverse component of magnetic
field $\langle B_n \rangle$. Similarly, the upstream Alfvén speed can be
determined from, e.g., $v_A = \hat{e}_{n} \cdot \left[ \vec{v}_A(\vec{p} -
\delta \hat{e}_{n}) + \vec{v}_A(\vec{p} + \delta \hat{e}_{n}) \right] / 2$,
which is necessary to estimate the Lundquist number $S$.

\item Finally, in order to estimate the length of the shear region, i.e. the
longitudinal dimension of the shear, we project the detector $S$ along the
vector parallel to the sheared component of the field, $S(s) = S(\vec{p} + s
\hat{e}_s)$, and analyze the decay of $S$ along $s$.  We measure the distance
$l$ between points where $S$ drops to a half of its maximum value in position
$\vec{p}$, treating $l$ as the longitudinal length of shear region.
\end{enumerate}

The procedure described above allows us to estimate the thickness $\delta$ and
the longitudinal dimension $l$ of all shear regions found in the domain.  For
each region we can determine the local direction of the shear $\hat{e}_s$ and
estimate other parameters required for determination of the local growth rate
$\gamma$, which are described more in the next subsection.

For illustration, in the left panel of Figure~\ref{fig:fit} we show a sketch of
a shear region (with arbitrary orientation) with vector field lines of the
opposite polarization (red and blue) with the local reference frame used to
project the field components on three axes $\hat{e}_s$, $\hat{e}_n$, and
$\hat{e}_g$, corresponding to the shear, transverse and guide components.  For
the case of magnetic shear, an extracted profiles of sheared (reconnecting)
$B_s(s)$, transverse $B_n(s)$, and guide $B_g(s)$ components (blue, orange, and
green respectively) along the direction normal to the current sheet are shown in
the right panel of Figure~\ref{fig:fit}.  The plot also shows the fitted
parameters $\delta B$, $B_{0}$ and $\delta$ of the sheared component (red dashed
line) together with the estimated stabilizing parameter $\xi$.

%
\subsection{Estimation of Conditions and Growth Rates for Tearing Instability}
\label{ssec:tr-tests}

The estimation of the growth rate of tearing mode in fluid simulations is not
trivial.  First of all, it is necessary to detect the locations of current
sheets, and once it is done, to estimate the local Lundquist number $S$ from the
upstream Alfv\'en speed $v_A$ and the thickness $\delta$ of each sheet region.
In order to take into account the effect of transverse field, we need to
estimate $\xi$ from the transverse and total magnetic fields, $B_n$ and $B$,
respectively.  Nevertheless, all there parameters are not sufficient enough
since we still need information about the local perturbations, which might be
especially difficult to characterize.  In the turbulent case, considered here,
we usually have a packet of waves of different amplitudes, wavenumbers and
directions traveling though the considered region.  First of all, we can
estimate the bounds for the perturbation wavenumber $k$ from the current sheet
region geometry.  Estimating the longitudinal length of the sheet, $l$, (see
step 7 in the previous section), we can determine the lower limit for the
wavenumber $k_{l} \approx 2 \pi / l$.  The upper wavenumber limit is determined
by the resolution of the simulation, i.e., $k_{h} \approx 2 \pi / h$. Therefore,
the permitted range of wavenumbers for local perturbations is $k_{l} \lesssim k
\lesssim k_{h}$.  For our analysis, it is enough, however, to consider $k_{max}$
corresponding to the maximum growth rate $\gamma_{max}$ assuming, that since we
have a turbulent region, there is a high probability finding fluctuations with
wavenumber equal to $k_{max}$ present in the region. Obviously, if $k_{max}$
lays out of the permitted range, the limited value of $k_{max} =
\min{(\max{(k_{max}, k_{l})}, k_{h})}$ should be considered in the estimation of
the tearing growth rate $\gamma_{max}$.  Also, we take into account that
$k_{max} \delta$ should be smaller than unity.

For the case of tearing instability, the algorithm described in
\S\ref{ssec:shear_detection} allows us to determine the thickness of local
current sheet $\delta$, the local upstream Alfv\'en speed $v_A$, the transverse
and total components of magnetic field, $B_n$ and $B$, respectively, and the
longitudinal scale of sheet region $l$.  From $\delta$ and $v_A$ we calculate
$S$ (using the explicit value of resistivity $\eta$ for which the simulation was
done), and from $B_n$ and $B$ we calculate $\xi$.  The estimated values of $S$
and $\xi$ are inserted into Eq.~(\ref{eq:tr_max_xi}), which is solved
numerically in order to get $p_{max}$.  Then, using  the left expression in
Eq.~(\ref{eq:tr_max}) we estimate $\alpha_{max}$ from $p_{max}$.  If
$\alpha_{max} = k_{max} \delta < k_{l} \delta$, however, its value is limited by
setting $\alpha_{max} = k_{l} \delta$ and estimating $p$ numerically directly
from Eq.~(\ref{eq:tr_grate_xi_full}) for a given $\alpha = \alpha_{max}$, using,
e.g., the Newton iterative root finder with the initial guess set $p_{max}$
obtained from Eq.~(\ref{eq:tr_max_xi}).  In this way we estimate the local
growth rate of tearing instability, $\gamma_{max} = p_{max} / \tau_R$, and the
corresponding wavenumber $k_{max}$.  The results of this analysis are presented
in \S\ref{sec:results:tearing} and \S\ref{sec:results:rates}.

%
\subsection{Estimation of Conditions and Growth Rates for Kelvin--Helmholtz Instability}
\label{ssec:kh-tests}

The Kelvin--Helmholtz instability is analyzed in a similar manner as the tearing
mode.  Here, we determine the positions of the velocity shear using the
algorithm described in \S\ref{ssec:shear_detection}.  Once a shear region is
detected, its thickness $\delta$ and longitudinal dimension $l$ are estimated.
These two parameters allow us to estimate the permitted range of the wavenumbers
of perturbation, $k_{l} \lesssim k \lesssim k_{h}$, where $k_{l}$ and $k_{h}$
are calculate in the same way as in the tearing instability, i.e., $k_{l} = 2
\pi / l$, and $k_{h} = 2 \pi / h$, however, $l$ refers here to the longitudinal
dimension of the velocity shear region.  From \ref{eq:kh-condition} we see that
the stability condition depends on the relative angle between the direction of
perturbation propagation $\hat{k} = \vec{k} / k$, the direction of local
magnetic field $\hat{B}_0 = \vec{B}_0 / B_0$, both with respect to the direction
of shear $\hat{U} = \vec{U} / U$.  Discussing the distribution of $U$ in
\S\ref{sec:results:kh} we consider two cases, $\vec{k} \parallel \vec{B}
\parallel \vec{U}$ and $\vec{k} \parallel \vec{B} \perp \vec{U}$.  However, in
the derivation of maximum growth rates described below we always assume the case
of three vectors parallel to each other.

\begin{figure*}[t]
\centering
\includegraphics[width=0.48\textwidth]{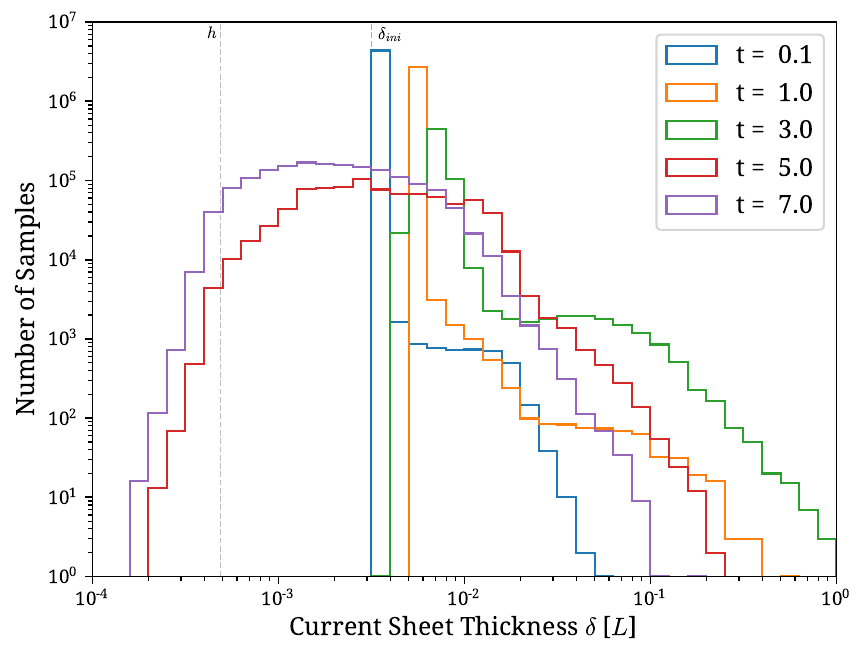}
\includegraphics[width=0.48\textwidth]{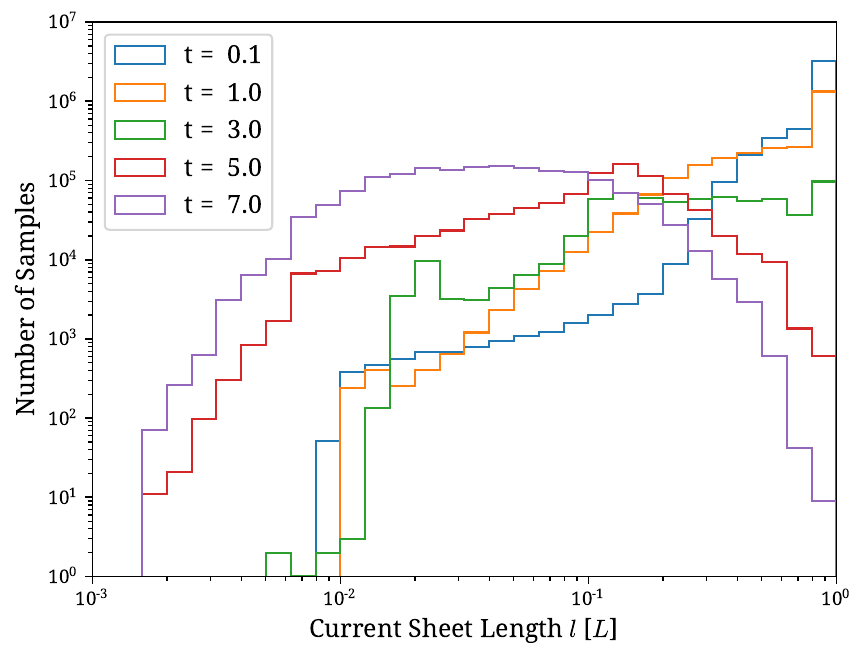}
\caption{Statistics of the current sheet thickness $\delta$ (left) and length
$l$ (right) for different evolution moments, $t = 0.1$, $1.0$, $3.0$, $5.0$,
$7.0$ (blue, orange, green, red, and purple, respectively).  The right vertical
dashed line (teal) shows the initial thickness of current sheet $\delta_{ini} =
3.16\times10^{-3}$, and the left line (grey) shows the effective grid size $h$.
\label{fig:csheet_dimensions}}
\end{figure*}

In order to determine the growth rate of the Kelvin--Helmholtz instability in
each detected region, we estimate the shear velocity $U = \delta v_s$ from the
interpolated transverse component of velocity (see step 5 in
\S\ref{ssec:shear_detection}).  We also estimate the upstream Alfvén speed $v_A$
in a similar way as in the tearing mode.  In this way we build a vector of
samples for the shear width $\delta$, the velocity shear amplitude $U$, and the
Alfvén speed $v_A$, necessary to verify the stability conditions and estimate
the growth rate from Eqs.~(\ref{eq:kh_grate_disc}), (\ref{eq:kh_grate_imhd}),
and (\ref{eq:kh_grate_mhd}).  In estimating the maximum growth rate
$\gamma_{max}$ in the case of discontinuous shear and incompressible MHD
(Eq.~\ref{eq:kh_grate_disc}) we estimate the wavenumber $k_{max}$ from the
condition $k_{max} = \max{(0.4 / \delta, k_l)}$.  The value 0.4 has been chosen
based on the $k_{max} \delta$ for the piecewise linear profile for the
hydrodynamic case (see Eq.~\ref{eq:kh_grate_hd} and the description below).  In
the case of MHD Kelvin-Helmholtz instability dispersion relations with the
piecewise linear profile (Eqs.~\ref{eq:kh_grate_imhd} and
\ref{eq:kh_grate_mhd}), both $\alpha_{max}$ and $\Omega_{max}$ have to be found
numerically using, for example, the golden section method \citep{Kiefer:1953}.
Once they are estimated, we calculate the maximum growth rate from $\gamma_{max}
= \operatorname{Im}(\Omega_{max}) \, U / \delta$ and $k_{max} = \alpha_{max} /
\delta$.  The results of the above described analysis of the Kelvin--Helmholtz
instability are presented in \S\ref{sec:results:kh} and \S\ref{sec:results:rates}.

\section{Analysis and Results}
\label{sec:results}

Before we describe our results, we should stress, that we present the
statistical results corresponding to the populations of cells which belong to
magnetic (current sheet) or velocity shear maximums and not individual shear
regions.  It means that we do not group cells as belonging to the same shear
region or not, or analyze any connectivity between these cells.  It also does
not mean, that we demonstrate the statistics of individual shear regions.  The
number of samples shown in, e.g., histograms, has only comparative, and not
absolute meaning, and since we analyze cells at shear maximums, it should not be
interpreted as a filling factor, once divided by the total effective number of
cells in the system.  The main objective of this work is to verify what
conditions are provided for the development of analyzed instabilities in the
reconnection-driven turbulence, and if they can be responsible for energy input
driving local turbulence.

%
\subsection{Tearing instability analysis}
\label{sec:results:tearing}

Before we estimate the growth rate of tearing instability, we shall analyze the
properties of current sheets in the system.  From
Eqs.~(\ref{eq:tr_grate_furth_short}) and (\ref{eq:tr_grate_furth_long}) we see,
that the growth rate $\gamma$ increases with the decrease of the current sheet
thickness $\delta$.  It also increases with the perturbation wavenumber $k$ for
$k < k_{max}$ (i.e., for long-wavelength modes), and decreases with $k$ for $k >
k_{max}$ (i.e., for short-wavelength modes).  These two quantities, $\delta$ and
$k$, also determine the instability condition $k \delta < 1$.  Therefore, we
will analyze them first.

\begin{figure*}[t]
\centering
\includegraphics[width=0.48\textwidth]{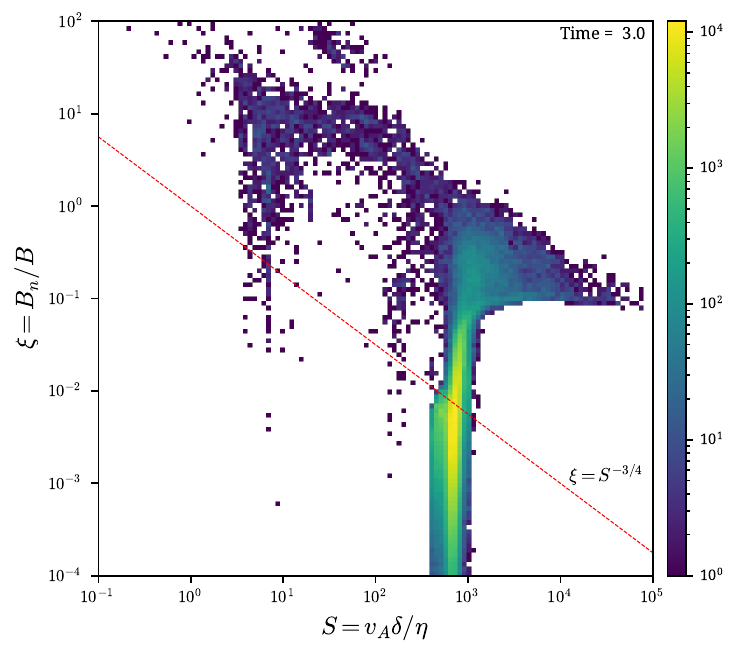}
\includegraphics[width=0.48\textwidth]{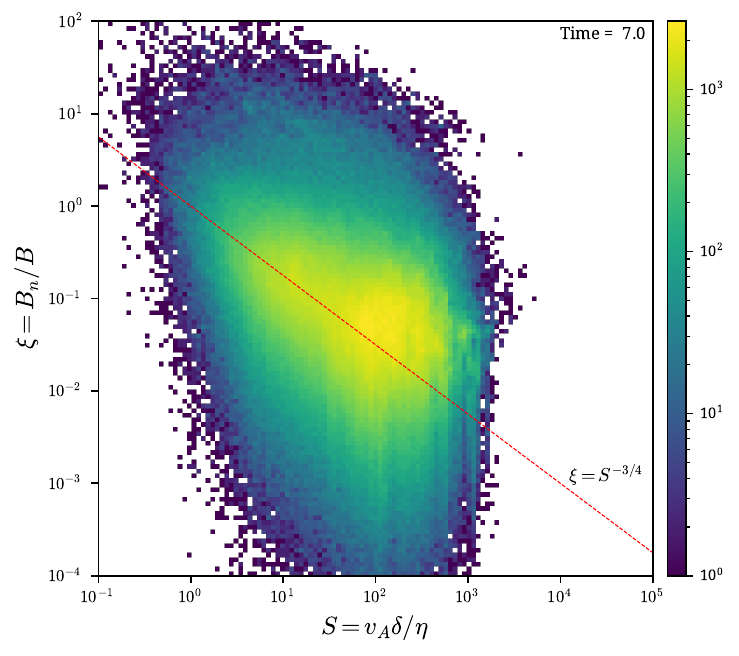}
\caption{Correlations between the ratio of the transverse component of magnetic
field to the magnetic field amplitude within the current density $\xi = B_n / B$
against the Lundquist number $S = \delta \, v_A / \eta$ at two evolution moments
$t = 3.0$ (left) and $t = 7.0$ (right).  Above the red dashed line the
transverse component starts to decrease the growth rate, eventually stabilizing
tearing mode.
\label{fig:correlations}}
\end{figure*}

In the left plot of Figure~\ref{fig:csheet_dimensions} we show the distribution
of current sheet thicknesses $\delta$ for all detected current sheet cells at
times $t = 0.1$, $1.0$, $3.0$, $5.0$, and $7.0$.  Two vertical lines correspond
to the effective cell size $h = 1/2048$ (left) and the initial current sheet
thickness $\delta_{ini} = 3.16\times10^{-3}$.  We see that at initial times two
populations of samples corresponding to current sheet regions form.  The first
dominating one is characterized by thickness which broadens to values several
times larger than the initial thickness $\delta_{ini}$, and the second one
characterized by very broad current sheets with $\delta$ comparable to a
fraction of the unit length $L$ (see, e.g., green line in
Fig.~\ref{fig:csheet_dimensions} corresponding to $t = 3.0$).  The population of
broad current sheets seems to be transient, since at $t = 5.0$ (red line in
Fig.~\ref{fig:csheet_dimensions}) it is significantly suppressed.  For $t \ge
5.0$, the distribution is not characterized by two populations anymore, and it
shifts to smaller values of $\delta$ with a significant fraction of the detected
current sheet samples comparable or below the effective cell size $h$,
indicating a sharp change of magnetic field orientation across the sheet (only
two cells to change the polarization of magnetic field lines) and probably
related to turbulent dynamics near the sheet plane.  On the other side, the
number of current sheet samples quickly decays with the value of $\delta$,
indicating that thick current sheets are not common in the system anymore.

Respectively, in the right plot of Figure~\ref{fig:csheet_dimensions} we show
the evolution of distribution of the lengths of current sheet regions at the
same times.  As expected, initially we have one current sheet plane, extended
over the whole box.  This is indicated by a significant number of samples of $l
\approx 1.0$ at $t = 0.1$ (blue line).  However, we can also see, that at this
very early time, a less significant population of samples with lengths being a
fraction of $L$ shows up.  We would interpret them as the points belonging to
parts of the current sheet already significantly deformed, since using our
analysis, we cannot determine if these points belong to the same or separated
current sheets.  It is important, however, that this population increases with
time, as seen at times $t = 1.0$ and $2.0$ (orange and green lines,
respectively).  At later times, $t > 3.0$, nearly all points belong to current
sheet regions with longitudinal dimension significantly shortened, comparing to
the box size, with values spread between $l = 10^{-2}$ and $10^{-1}$ at $t =
7.0$ (purple line).

\begin{figure*}[t]
\centering
\includegraphics[width=0.48\textwidth]{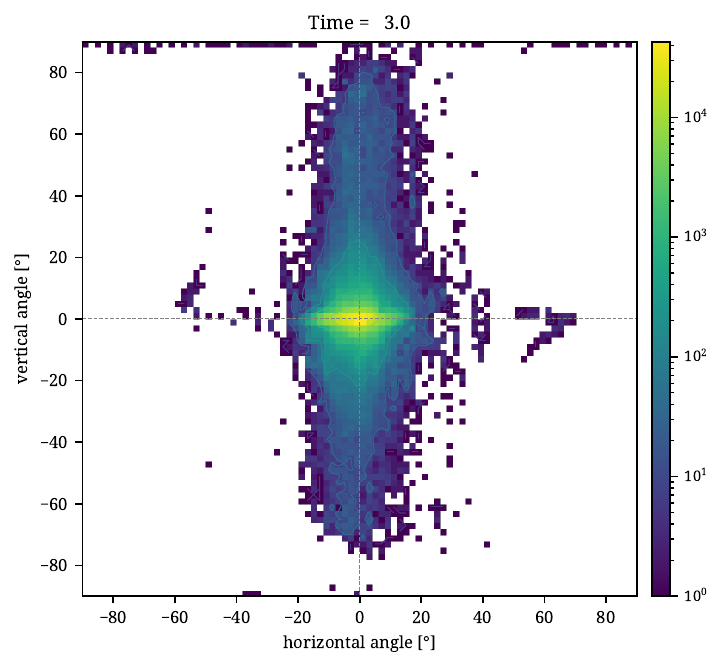}
\includegraphics[width=0.48\textwidth]{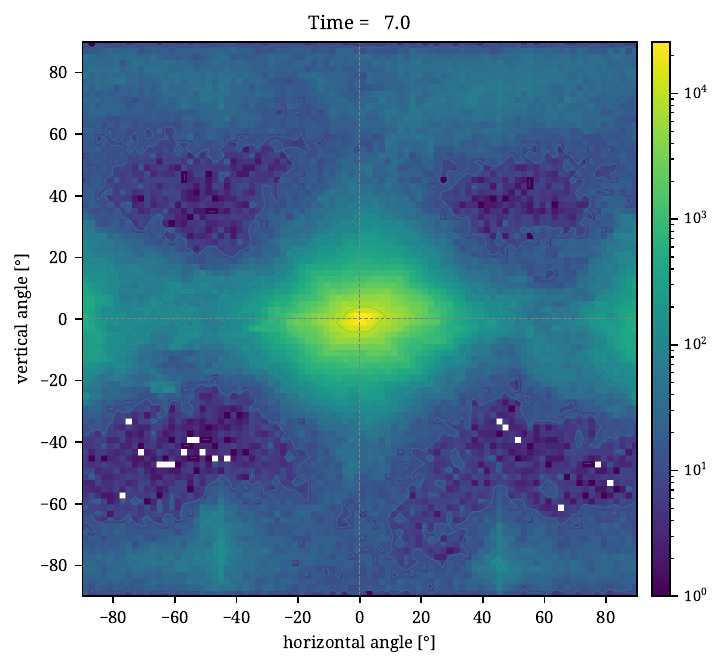}
\caption{Distribution of the magnetic shear directions in unstable cells
($\gamma_{max} > 10^{-3}$) for the same model and moments as shown in
Fig.~\ref{fig:correlations}.  The horizontal angle corresponds to the azimuthal
angle projected on the XZ plane with respect to the X axis.  The vertical angle
is the angle between the shear direction and the XZ plane.
\label{fig:tr_angular}}
\end{figure*}

Analyzing Figure~\ref{fig:csheet_dimensions} we can deduce, that at initial
times tearing mode should be a preferential instability for fluctuations
generation, since the current sheets are characterized by relatively thin and
extended current sheets with $\delta \approx 0.005 - 0.05$ and $l \lesssim 1.0$.
Since $k \delta < 1.0$ for instability, it indicates that the limit for allowed
wavenumbers is $k \lesssim 200$.  Recalling from \S\ref{ssec:simulations} the
imposed perturbation wavenumber $k = 64 \pi \approx 201$ at $t = 0.0$ indicates
that the system should be nearly stable initially.  At the final times $\delta$
decreases to values $\delta \approx 0.0002 - 0.1$, which should significantly
enhance the development of tearing mode, however, the fragmentation or
deformations of the current sheet decrease significantly the length $l$ of the
current sheet regions impeding the development of long wave perturbations.
Still, according to the instability condition $k \delta < 1$, perturbations with
wavenumbers up to $k \approx 5000$ could be be unstable.

The analysis above did not give a clear response to the question, if turbulence
could be generated by tearing mode initially.  At later times, the turbulence
develops in regions where current sheet is thinner, potentially increasing the
growth rate of the instability.  At the same time, however, it is possible that
the same turbulence generates component of magnetic field normal to current
sheet, which, according to Eq.~(\ref{eq:tr_grate_xi_full}), may suppress the
instability.  In order to analyze the stabilizing effect of this component, we
show the correlations between the normalized transverse component of magnetic
field $\xi = B_n/B$ and the Lundquist number $S = v_A \delta / \eta$ in
Figure~\ref{fig:correlations} for two moments, $t = 3.0$ (left panel) and $t =
7.0$ (right panel).  The red line, corresponding to the relation $\xi =
S^{-3/4}$, divides the plot into two regions: one below the line, where $\xi$
has negligible effect, and another above the line, where the stabilization by
$\xi$ starts to be significant and increases with the distance from the red
line.  In the left panel ($t = 3.0$) we see, that $\xi$ is not important for
most of the detected cells with their $S$ values concentrated slightly below
$10^3$ and the parameter $\xi$ spreading up to value $10^{-2}$.  A partial
stabilization in the upper tail, i.e. for $\xi \approx 10^{-2} - 10^{-1}$,
already takes place.  We notice, that it has a characteristic increase in the
direction of larger values of $S$ at $\xi \approx 10^{-1}$.  This is probably
due to the broadening of the current sheet seen in the left panel of
Figure~\ref{fig:csheet_dimensions}.  The points above $\xi = 1.0$, although
completely stabilized by $\xi$, are statistically insignificant.

At later time, $t = 7.0$, shown in the right panel of
Figure~\ref{fig:correlations}, the situation is very different.  The points of
distribution spread toward lower values of $S$, roughly between $10^0 and 10^3$,
and across many orders of magnitude along the stabilization parameter $\xi$,
nearly up to $10^2$.  We see a significant concentration of detected samples
somewhat above the red line dividing two regions of $\xi$ importance.  Moreover,
the points in the range $10^3 < S < 10^5$ disappeared nearly completely.  The
spread to the left along the horizontal direction should be attributed to the
decrease of current sheet thicknesses due to the action of turbulence, which is
also responsible for the generation of the transverse component $B_n$.  From the
distributions for two different times shown in Figure~\ref{fig:correlations} we
see that the generation of $\xi$ by turbulence cannot be ignored in the
estimation of the growth rate of tearing instability.

An interesting question to ask is what is the principal direction of magnetic
shear in the detected current sheet cells at different moments, considering the
presence of a guide field and weak initial perturbations.  In
Figure~\ref{fig:tr_angular} we show the angular distribution of the shear
direction only for the unstable cells (for which $\gamma_{max} > 10^{-3}$) at
two moments, $t = 3.0$ (left panel) and $7.0$ (right panel).  We see that at
$t=3.0$ the shear direction is still strongly concentrated along the
X-direction, the direction of the reconnecting component, spreading roughly from
$-20^\circ$ to $20^\circ$ in the azimuthal and from $-10^\circ$ to $10^\circ$ in
the vertical directions, with some very rare events reaching even higher
altitudes.  At the final time, $t = 7.0$, however, the distribution of
directions, although still strongly concentrated along the X axis, characterizes
by a much larger spread in both directions.  This indicates, that the turbulence
acting on the current sheet can significantly bend it, modifying its local
topology.

%
\subsection{Kelvin--Helmholtz instability analysis}
\label{sec:results:kh}

\begin{figure*}[t]
\centering
\includegraphics[width=0.48\textwidth]{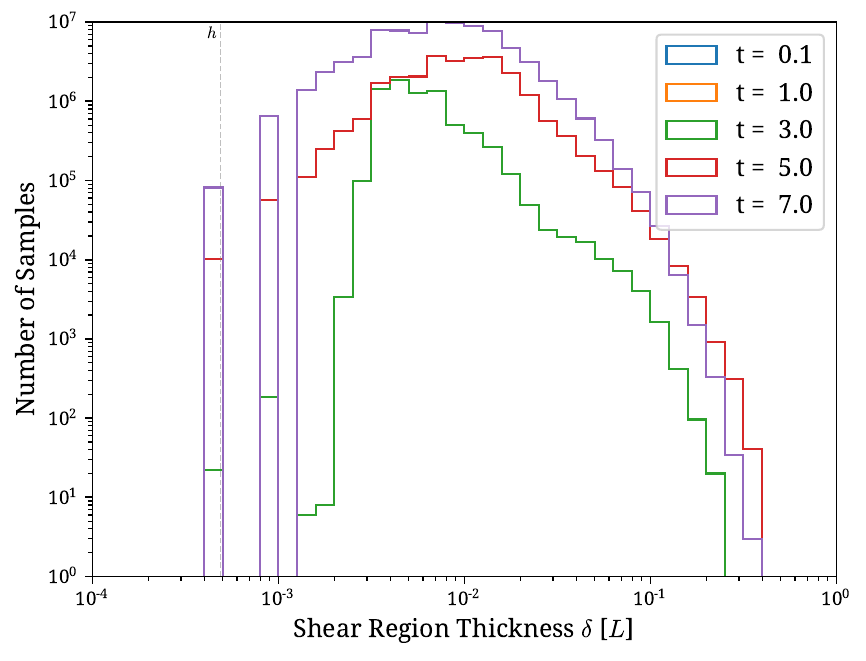}
\includegraphics[width=0.48\textwidth]{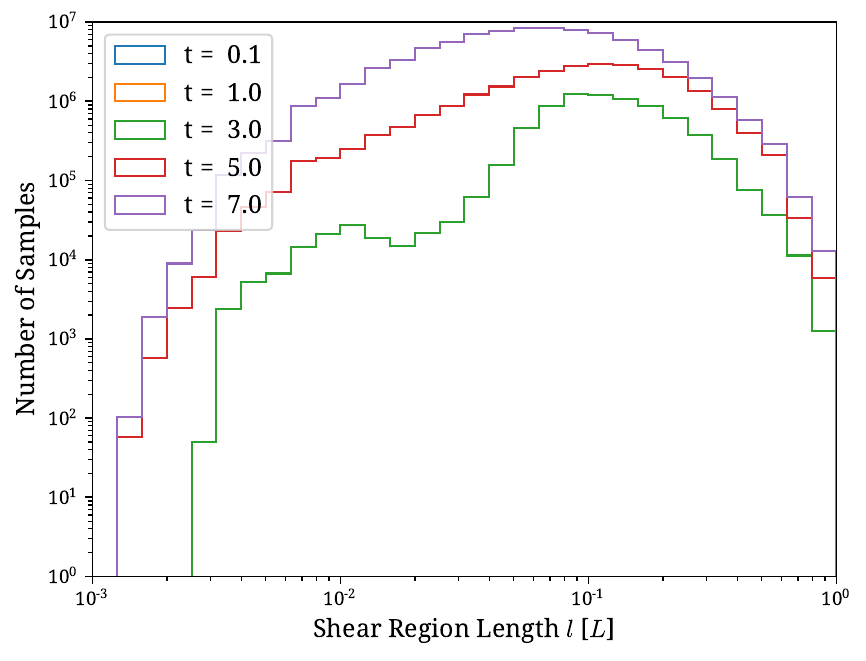}
\caption{Statistics of the velocity shear region thickness $\delta$ (left) and
length $l$ (right) at different evolution moments, $t = 0.1$, $1.0$, $3.0$,
$5.0$, $7.0$ (blue, orange, green, red, and purple, respectively).  The vertical
dashed line (grey) shows the effective grid size $h$.
\label{fig:kh_shear_dimensions}}
\end{figure*}

\begin{figure*}[t]
\centering
\includegraphics[width=0.48\textwidth]{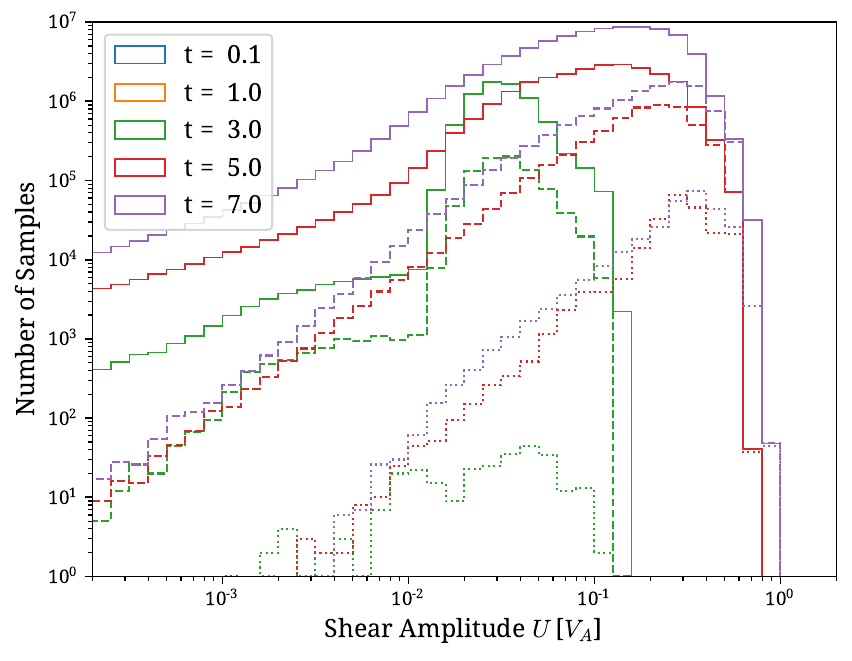}
\includegraphics[width=0.48\textwidth]{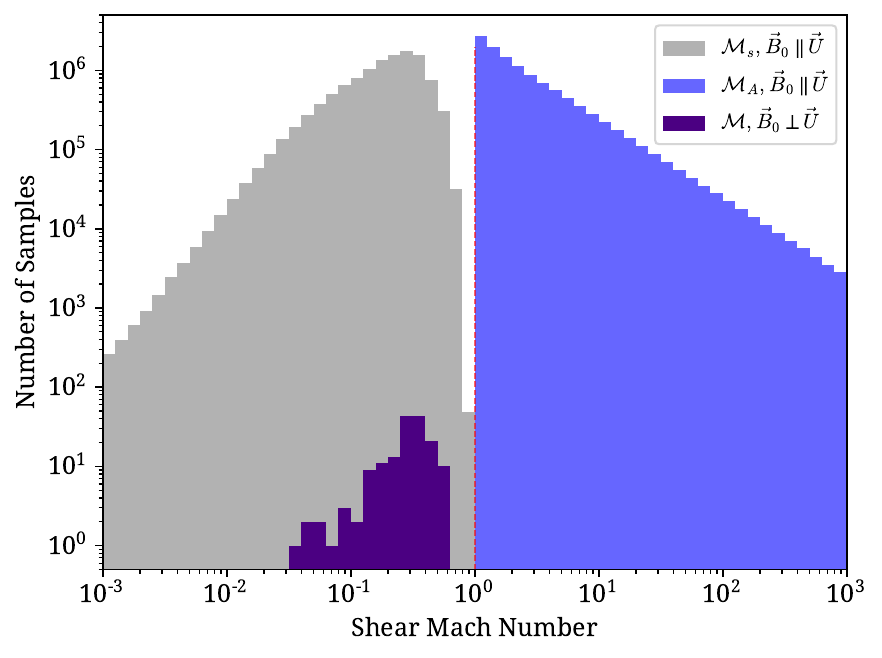}
\caption{{\it Left:} Evolution of the distribution of the velocity shear
strength $U$ for different times.  Solid lines correspond to all cells where
shear was detected, while dashed and dotted lines correspond to the cells which
are the Kelvin--Helmholtz unstable under assumption of the perturbation
wavevector $\vec{k}$ forming an angle of $0^\circ$ and $75^\circ$ with the shear
direction, respectively.
{\it Right:} Distribution of sonic ${\cal M}_s = U/a$ (grey), Alfvénic ${\cal
M}_A = U/v_A$ (blue) and magnetosonic ${\cal M}_s = U/c_M$ (indigo) Mach numbers
in the Kelvin--Helmholtz unstable samples at $t = 7.0$.  As predicted by
\cite{MiuraPritchett:1982}, if assuming $\vec{B}_0 \parallel \vec{U}$, ${\cal
M}_s < 1$ and ${\cal M}_A > 1$ for all unstable cells, while for $\vec{B}_0
\perp \vec{U}$, ${\cal M} < 1$.  The red dashed line corresponds to Mach number
equal $1.0$.
\label{fig:kh-mach}}
\end{figure*}

Similarly to the tearing mode analysis, we start by showing the distributions of
the thickness $\delta$ and length $l$ of the velocity shear regions in
Figure~\ref{fig:kh_shear_dimensions}.  The first interesting observation is that
there are no detected velocity shear regions for times $t = 0.1$ and $1.0$, or
the shear strength is too weak, below the threshold value $U_{min} = 10^{-4}$
set in the shear detection algorithm.  These distributions are of all cells at
detected velocity shear maximums.  At $t = 3.0$ we already see a number of cells
belonging to shear regions of thicknesses $\delta$ between $2 \times 10^{-3}$
and about a half of length unit, with distributions peaking at values below
$10^{-2}$ for $t = 3.0$ and values larger than $10^{-2}$ at $t = 7.0$. Looking
at the right panel of Figure~\ref{fig:kh_shear_dimensions} we see that the
longitudinal dimensions of these shear regions spread from several cells to
unity, indicating a generation of nearly global shear in the computational
domain.  Transforming these lengths into wavenumbers indicates, that
perturbations of any $k$, from $k = 2 \pi$ up to nearly $k \sim 5000$, may grow
due to Kelvin--Helmholtz instability, if appropriate conditions are fulfilled in
the local shear region.  The peak value for the longitudinal dimension of shear
regions is about $l \sim 0.1$, decreasing slightly for later times,
corresponding to the wavenumbers of $k \sim 60 - 100$.

\begin{figure*}[t]
\centering
\includegraphics[width=0.48\textwidth]{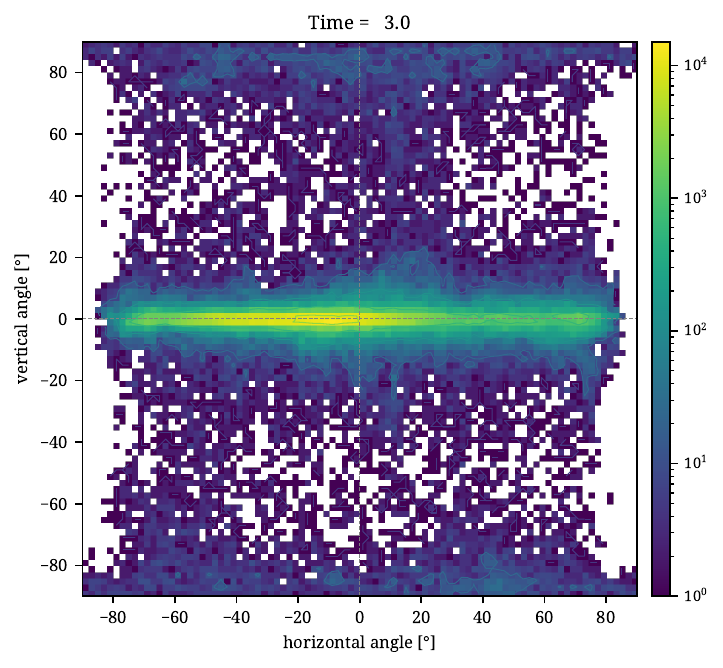}
\includegraphics[width=0.48\textwidth]{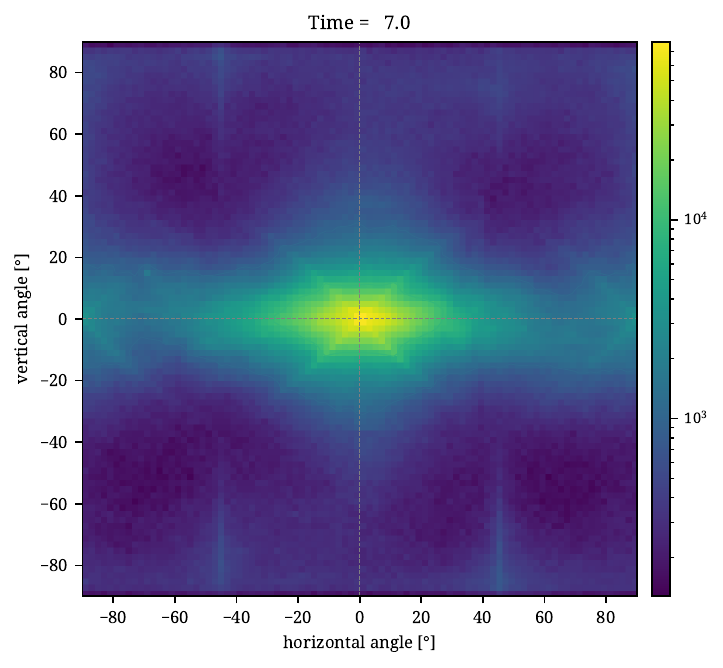}
\caption{Distribution of the velocity shear direction in unstable cells for the
same moments as shown in Fig.~\ref{fig:correlations}.  The horizontal angle
corresponds to the azimuthal angle projected on the XZ plane with respect to the
X axis.  The vertical angle is the angle between the shear direction and the XZ
plane.
\label{fig:kh_angular}}
\end{figure*}

The most important parameter in the development of the Kelvin--Helmholtz
instability is the shear amplitude $U$.  In the left panel of
Figure~\ref{fig:kh-mach} we show the evolution of distribution of $U$ for all
cells where the maximum of shear was detected (solid lines) and only for cells
which are unstable with assumption that the angle between the wavevector
$\vec{k}$ and the shear direction is $0^\circ$ (dashed) and $75^\circ$ (dotted),
according to the sufficient condition for the Kelvin--Helmholtz instability
(Eq.~\ref{eq:kh-condition}).  We notice, that although the shear is relatively
common after $t = 3.0$, only cells with the strongest $U$ are in fact unstable
(i.e., the corresponding growth rates determined from Eq.~\ref{eq:kh_grate_mhd}
are larger than zero).  We see that for these unstable cells the shear strength
spreads between $10^{-2}$ to nearly $1.0$, measured in Alfvén speed $v_A$.  At
later times, the distribution peaks at values close to $v_A$.  This plot clearly
indicates, that strong shear can be generated in such systems in relatively
short time.

In the right panel of Figure~\ref{fig:kh-mach} we verify prediction for the
compressible system by \cite{MiuraPritchett:1982}, stating that if the magnetic
field is parallel to the shear ($\vec{B}_0 \parallel \vec{U}$), if the sonic
Mach number ${\cal M}_s > 1.0$ or ${\cal M}_A < 1.0$ the instability is
stabilized.  We show distributions of both Mach numbers for the last snapshot of
our simulation, at $t = 7.0$.  Clearly, all unstable cells have sonic Mach
number ${\cal M}_s < 1.0$ and Alfv\'enic Mach number ${\cal M}_A > 1.0$, being
in a perfect agreement with the theoretical prediction.  Also, if assuming a
perpendicular magnetic field ($\vec{B}_0 \perp \vec{U}$), ${\cal M} < 1.0$ for
all unstable cells.  As we can see in the right panel of
Figure~\ref{fig:kh-mach}, this case agrees with the predictions too.

Similarly to tearing mode analysis, we show the distribution of velocity shear
directions in Figure~\ref{fig:kh_angular} for two moments: at $t = 3.0$ (left
panel), when the Kelvin--Helmholtz unstable cells start to appear, and at the
final moment $t = 7.0$ (right panel), when the instability is already developed.
 We see that at $t = 3.0$ the Kelvin--Helmholtz instability can operate only in
the plane extended significantly in the horizontal direction from $-80^\circ$ to
$80^\circ$ and by a few degrees in the vertical one with respect to the X
direction.  There is also observed a small, statistically insignificant
distribution of shear direction in the direction perpendicular to the shear
plane.  At the final time ($t = 7.0$), the velocity shear directions are
scattered over all angles in both directions, azimuthal and vertical, although
the most statistically significant part is within $20^\circ$ from the X axis,
extended somewhat more in the azimuthal direction across all angles.

%
\subsection{Evolution of the Growth Rates: Tearing vs. Kelvin--Helmholtz}
\label{sec:results:rates}

\begin{figure*}[t]
\centering
\includegraphics[width=0.47\textwidth]{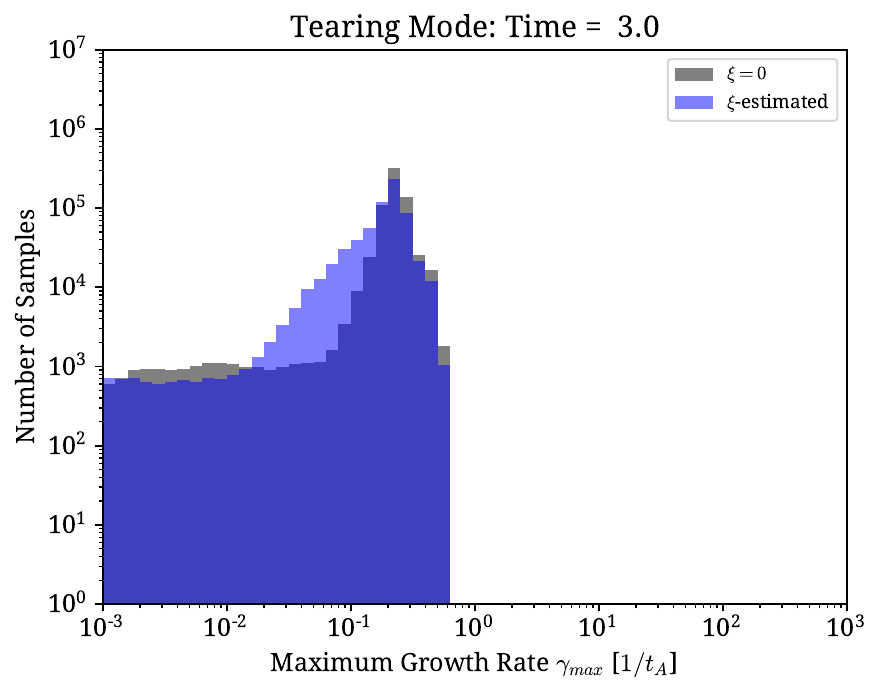}
\includegraphics[width=0.47\textwidth]{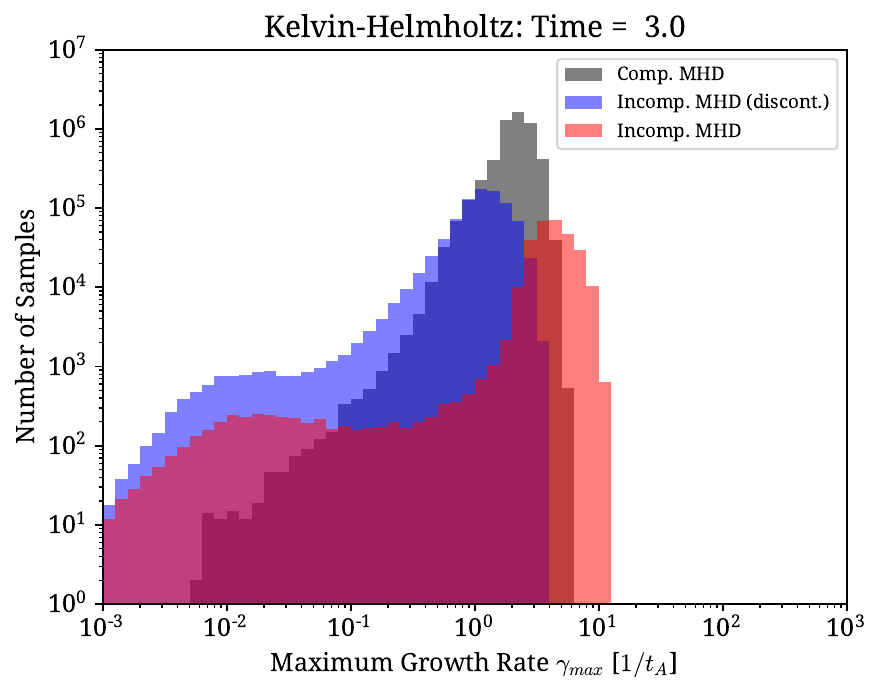}
\includegraphics[width=0.47\textwidth]{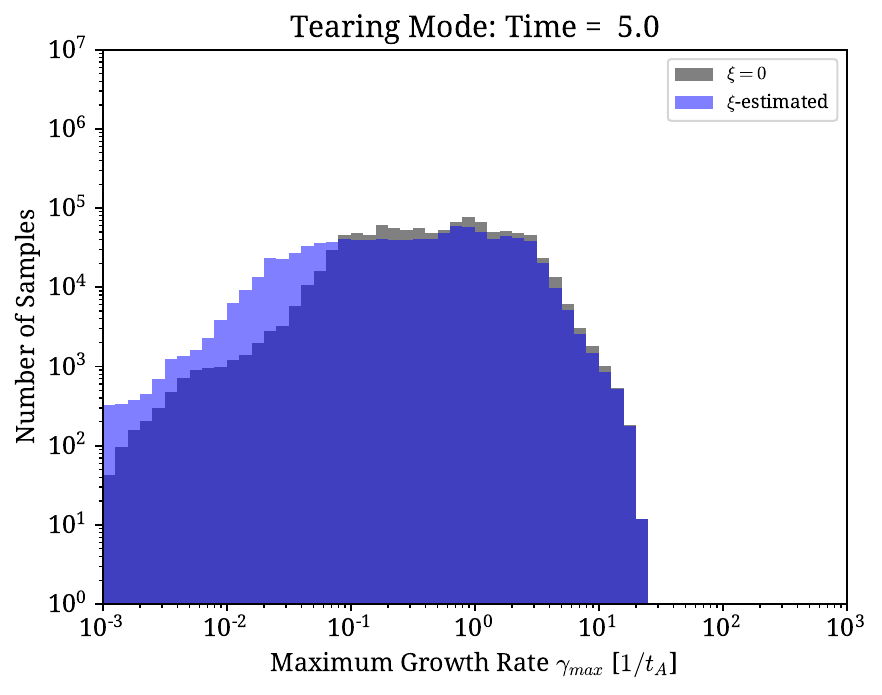}
\includegraphics[width=0.47\textwidth]{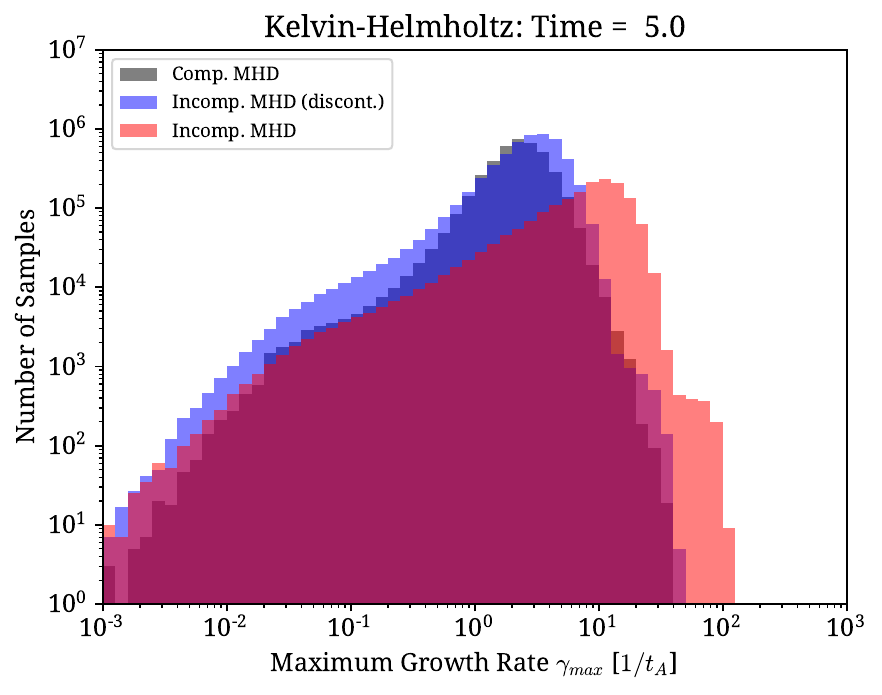}
\includegraphics[width=0.47\textwidth]{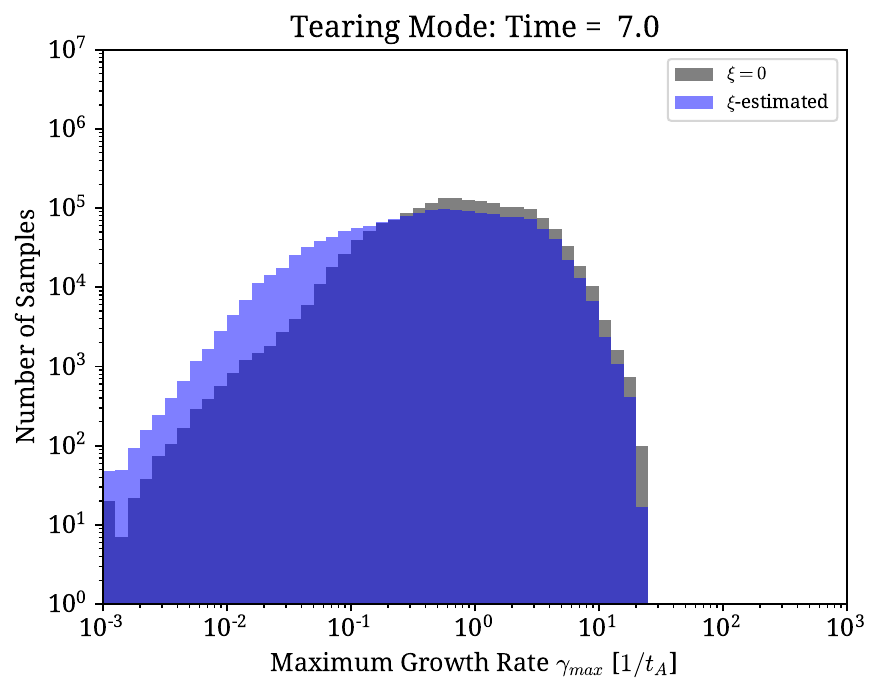}
\includegraphics[width=0.47\textwidth]{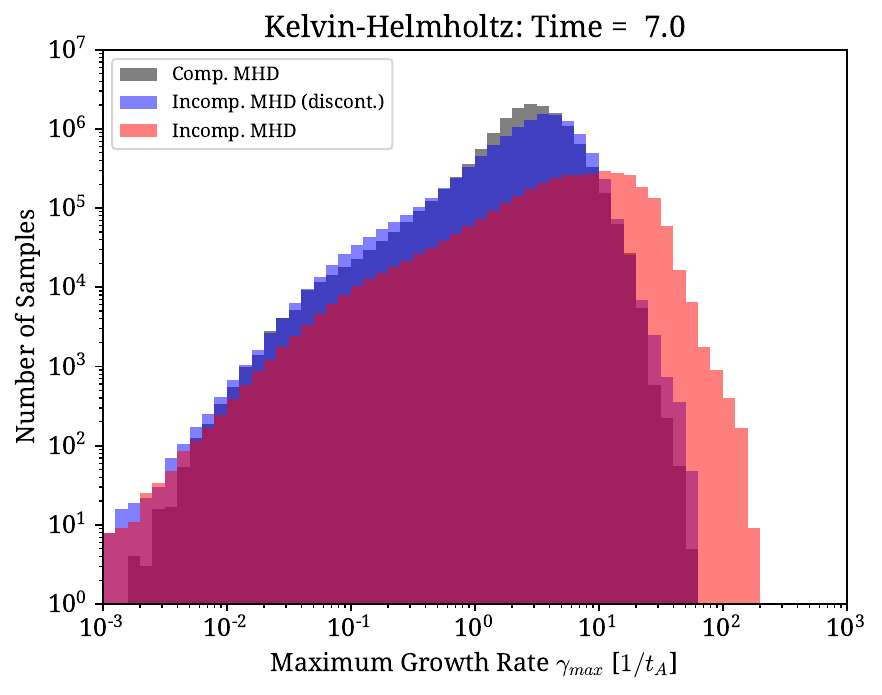}
\caption{Evolution of the distribution of maximum growth rates for tearing mode
(left) and Kelvin--Helmholtz (right) at three different times, $t = 3.0$, $5.0$,
and $7.0$ (upper, middle, and bottom, respectively).  For tearing mode, the grey
and blue distributions correspond to $\gamma_{max}$ estimated from
Eq.~(\ref{eq:tr_max_xi}), with $\xi$ set to zero and to value estimated from
simulations, respectively.  In the $\xi$-estimated distribution, only cells
which fulfilled the condition $\xi < \alpha_{max}$ were taken into account.  For
the Kelvin--Kelmholtz instability, the blue, red, and grey distributions
correspond to $\gamma_{max}$ estimated from Eqs.~(\ref{eq:kh_grate_disc}),
(\ref{eq:kh_grate_imhd}), and (\ref{eq:kh_grate_mhd}), respectively, with the
respective validity conditions taken into account.
\label{fig:grate-evol}}
\end{figure*}

Supported by the results from previous subsections, which show the analysis of
the factors important for the development of tearing and Kelvin--Helmholtz
instabilities, we can now compare the estimated maximum growth rates for both
instabilities.  This will help us to determine which one dominates.  The way we
estimate $\gamma_{max}$ (and corresponding $k_{max}$) was already described in
details in \S\ref{ssec:tr-tests} and \S\ref{ssec:kh-tests} for tearing and
Kelvin--Helmholtz instabilities, respectively.  As we already have shown, the
range of possible wavenumbers of perturbations, $k$, for both instabilities can
be estimated from the longitudinal dimensions of individual shear regions, $l$,
and the effective cell size, $h$.  Of course, if the estimated $k_{max}$ lays
out of the allowed range, $k_l < k < k_h$, it is limited to $k_{max} =
\min{(\max{(k_{max}, k_{l})}, k_{h})}$ and a new $\gamma_{max}$ is calculated
accordingly.  It is also important to mention, that since the dispersion
relations provided in \S\ref{sec:instabilities} were derived under certain
assumptions, i.e., Eq.~(\ref{eq:tr_max_xi}) is valid for $\xi < \alpha$ for
tearing mode and Eqs.~(\ref{eq:kh_grate_imhd}) and (\ref{eq:kh_grate_mhd})
(incompressible and compressible MHD, respectively) are valid for ${\cal M}_A^2
\gg 1$ and $\mu^2 \ll 1$, respectively, for the Kelvin--Helmholtz instability,
only cells which fulfilled them were processed to obtain the results presented
here.

In Figure~\ref{fig:grate-evol} we show the distributions of estimated maximum
growth rates for both instabilities.  The statistics for tearing mode and
Kelvin--Helmholtz instability are shown in the left and right column,
respectively.  Three different time moments were chosen, $t = 3.0$, $5.0$, and
$7.0$, shown in the upper, middle, and lower rows, respectively.  We see, that
at earlier times (upper left panel), the tearing mode, even though expected to
be a dominant one, has relatively low maximum growth rates, $\gamma_{max} <
0.7$, with the distribution peaking at $\gamma_{max} \approx 0.2 - 0.3$.  We
should mention that considering the initial setup, the maximum growth rate was
estimated to a value $\sim 0.7$ at the initial current sheet.  Therefore, we see
a small decrease in $\gamma_{max}$ at early stages.  Nevertheless, the number of
maximum magnetic shear cells increases at later time with the distribution of
$\gamma_{max}$ reaching values up to $30.0$ and most of the cells characterized
by $\gamma_{max} > 0.1$ (see the middle and lower left panels in
Fig.~\ref{fig:grate-evol}).  We should note that the distributions of the
maximum growth rates for tearing instability are not too sensitive for the
traverse component, represented by the parameter $\xi$.  Even at the final time
of simulation, $t = 7.0$ (the lower left panel in Fig.~\ref{fig:grate-evol}),
when the turbulence is already developed in the initial current sheet region,
the growth rate $\gamma_{max}$ for the estimated value of $\xi$ is reduced only
in a small fraction of cells.

The Kelvin--Helmholtz instability is relatively negligible initially (see the
$U$ distribution in the left plot in Fig.~\ref{fig:kh-mach} for $t < 30.0$).  At
$t = 30$ (the upper right plot in Fig.~\ref{fig:grate-evol}) the velocity shear
regions have already developed and can initiate the turbulence production
through the instability.  As seen in the right column of
Figure~\ref{fig:grate-evol} the maximum values of $\gamma_{max}$ reach $4-10$
depending on the used dispersion relation, with the number of cells comparable
(blue and red) for incompressible MHD or even larger (grey) for compressible
MHD.  Taking into account the compressibility decreases $\gamma_{max}$ and
stabilizes the low shear regions ($U < 0.1$), as expected.   However, even
though the tearing mode is widespread, its maximum growth rates are smaller to
those related to Kelvin--Helmholtz instability at later times.  See the middle
and lower panel in Figure~\ref{fig:grate-evol} were the $\gamma_{max}$
distributions for Kelvin--Helmholtz somewhat dominate those for tearing mode,
both in terms of the reached growth rates and the number of detected cells.

At the final moment ($t = 7.0$), the distributions of growth rates for
Kelvin--Helmholtz instability when applying the compressible MHD derivation of
dispersion relation with a smooth profile and the incompressible MHD with the
discontinuous profile (Eqs.~\ref{eq:kh_grate_disc} and \ref{eq:kh_grate_mhd},
respectively) are very similar.  Once the compressibility is ignored and a
smooth shear profile considered, the instability can reach values up to nearly
$200$.  If we recall the distribution of Mach numbers shown in
Figure~\ref{fig:kh-mach}, all unstable cells are characterized by low sonic Mach
numbers, indicating that indeed $\gamma_{max}$ could be enhanced by a factor of
few.

\section{Discussion}
\label{sec:discussion}

%
\subsection{Limitations of our approach}

Our approach in analyzing tearing and Kelvin-Helmholtz instabilities is robust,
however, it has it drawbacks, which should be pointed out.  First of all, in the
presence of growing turbulent fluctuations, it is nearly impossible to determine
the characteristics of the perturbations present in the analyzed local shear
region.  In order to determine the growth rate precisely, we would have to
posses information about the wavenumber and direction of each local
perturbation.  In order to compensate the lack of these data, we determine the
plausible range of wavenumbers for which the instabilities would have been
unstable.  Therefore, they are limited to the range $k \in (k_{l}, k_{h})$ as
described in \S\ref{ssec:tr-tests} for tearing and \S\ref{ssec:kh-tests} for
Kelvin--Helmholtz instabilities.  In the derivation of the maximum growth rates
$\gamma_{max}$ we determine the corresponding perturbation wavenumber $k_{max}$.
If this wavenumber lays out of the range $(k_{l}, k_{h})$, it is limited and the
growth rate is calculated directly from the dispersion relations.  Moreover, we
apply all the conditions related to stability and the validity of dispersion
relations.  However, the condition for the stability of the Kelvin--Helmholtz
depends on the direction of perturbation and upstream magnetic field.  In the
derivation of $\gamma_{max}$ we assume $\vec{k} \parallel \vec{U}$, using the
fact that the initial perturbation was isotropic, and once the turbulence is
developed it is possible to find modes parallel to the shear.

As we described in the beginning of \S\ref{sec:results} our results are based on
the statistics extracted from the cell by cell analysis and the data were
collected only for cells belonging to shear maximums.  It means that the
measurement do not represent the individual shear regions and for which the
analysis is done separately.  For example, in the case of tearing instability
analysis, we have one current sheet initially crossing the whole computational
box, therefore the points represent the cells along the plane of the maximum of
current density.  Due to the developing turbulence, this current sheet is being
deformed and eventually broken into a number of current sheet regions, not
necessarily separated, but interlinked in a complex manner.  Therefore, our
analysis should be understood as a comparative one between two instabilities.
This should be kept in mind especially when interpreting the statistics of
longitudinal dimensions of shear regions.

The derivation of analytical dispersion relations for tearing and
Kelvin--Helmholtz instabilities requires significant simplifications.  Usually a
two uniform regions of density, pressure, magnetic field and velocity are
considered connected by transition layer.  This layer can be discontinuous or
smooth.  The smooth case is usually approximated by a piecewise linear or
hyperbolic tangent profile.  The dispersion relations are different for these
three different profiles, giving different dependence of the growth rate
$\gamma$ on the perturbation wavenumber $k$ \cite[see,
e.g.,][]{Furth_etal:1963, SomovVerneta:1989, Chandrasekhar:1961,
OngRoderick:1972, Walker:1981, MiuraPritchett:1982, Chen_etal:1997,
BerlokPfrommer:2019}.  For a discontinuous profile ($\delta \approx 0$) the is
no limit on the wavenumber.  The condition $k \delta < 1$ is always fulfilled in
the case of tearing mode, and the growth rate of Kelvin--Helmholtz instability
depends linearly on $k$, resulting in arbitrary large values of $\gamma$. Taking
into account a smooth transition within the shear region results in the
existence of the wavenumber $k_{max}$ for which the growth rate has its maximum
$\gamma_{max}$, and which typically is related to the thickness of the region $k
\delta \approx 1$.  This maximum growth rate is usually a fraction of the growth
rate corresponding to the discontinuous shear.  It is relatively difficult to
determine the exact profile in the shear regions detected in our simulation.
Specially in the case of broad shear and in the presence of turbulence the
profile can be very different from piecewise linear or hyperbolic tangent ones.
Nevertheless, after determining the thickness $\delta$ and upstream fields,
$\vec{B}_{0}$ and $\vec{U}$, we assumed they represent the finite width profiles
used in derivation of the corresponding to each instability dispersion relation.
Moreover, in the calculation of tearing growth rates we did not take into
account the compressibility, although its effect should be relatively
negligible, since the velocity fluctuations do not reach velocities larger than
the sound speed $a$.

%
\subsection{Turbulent reconnection as a dominant process}

Suggested 20 years ago, the turbulent reconnection model has gotten significant
support both from subsequent numerical \cite[see][]{Kowal_etal:2009,
Kowal_etal:2012, Kowal_etal:2017, Eyink_etal:2013, Oishi_etal:2015,
Takamoto_etal:2015, Beresnyak:2017, Takamoto:2018}, theoretical
\citep{Eyink_etal:2011, Eyink:2011, Eyink:2015, Lazarian_etal:2015,
Lazarian_etal:2019}, as well as  observational
\cite[see][]{CiaravellaRaymond:2008, Sych_etal:2009, Sych_etal:2015,
KhabarovaObridko:2012, Lazarian_etal:2012, Santos-Lima_etal:2013,
Leao_etal:2013, Gonzales-Casanova_etal:2018} studies.  At the moment of its
introduction the model was an alternative to the Hall-MHD models predicting
Petschek X-point geometry of reconnection point, i.e., very regular type of
reconnection.  The later model required plasma to be collisionless, which is in
contrast to the turbulent one, which did not depend on any  plasma microphysics
and was applicable to both collisional and collisionless media. It was later
understood that the X-point geometry is not tenable in realistic settings.
Instead, the tearing reconnection \cite[see][]{Syrovatskii:1981,
Loureiro_etal:2007, Bhattacharjee_etal:2009} became the main alternative
scenario for the turbulent model.  So far 2-dimensional simulations demonstrated
fast reconnection for both MHD and kinetic regimes. Compared to earlier Hall-MHD
reconnection that necessarily required collisionless plasma condition this was
definitely an important improvement. The tearing reconnection shares many
features with the turbulent one.  For instance, Hall-MHD Petschek-type
reconnection, due to the presence of slow shock crossing the boundary, required
a particular set of boundary conditions that was difficult to preserve in the
realistic setting with the random external perturbations
\cite[see, e.g.][]{Forbes:2001}.  In the presence of random/turbulent motions,
inevitable in astrophysical settings, the Petscheck solution would not be
satisfied.  The tearing plasmoid reconnection is a more robust scheme therefore.

With two reconnection processes providing fast reconnection, it is important to
understand the applicability of each.  It has been numerically demonstrated in
\cite{Kowal_etal:2009} that including additional microscopic effects simulating
enhanced plasma resistivity does not change the turbulent reconnection rate.
This agrees well with the theoretical expectations in turbulent reconnection
\cite[see LV99 and][]{Eyink:2011}, in particular with the generalized Ohm's law
derived in \cite{Eyink:2015}.  As a result, if media is already turbulent, one
does not expect to see effects of tearing reconnection.  With the existing
observational evidence about the turbulence of astrophysical fluids this means
that the turbulent reconnection is dominant for most of the cases.  For
instance, we expect the turbulent reconnection to govern violation of the flux
freezing in turbulent fluids.  This results in reconnection diffusion that
governs star formation \citep{Lazarian_etal:2012}, induces the violations of the
structure of the heliospheric current sheet and the Parker spiral
\citep{Eyink:2015}.

The numerical results on flux freezing violation that follows from the LV99
theory cannot be possibly explained with the tearing reconnection.  This clearly
demonstrates that there are situations when the turbulent reconnection is at
work, while tearing reconnection is not expected.

The “pure” problem of self-driven turbulent reconnection was the focus of our
study in \cite{Kowal_etal:2017}.  There we showed that in the absence of the
external turbulence driving the turbulence develops in the reconnection region
and this turbulence has the properties corresponding to the expectations of the
MHD turbulence.  This was in contrast to \cite{HuangBhattacharjee:2016} who
claimed that turbulence produced in reconnection regions is radically different
from the \cite{GoldreichSridhar:1995} one.  The properties of turbulence are
important, as the LV99 magnetic reconnection and closely connected to it
Richardson dispersion \citep{Eyink_etal:2011} are proven to work in conditions
where no tearing instability is expected.  Therefore, if such type of turbulence
is present in the reconnection regions it is expected to induce fast
reconnection.  The correspondence of the reconnection rates in self-driven
reconnection with the expectations of the LV99 theory was demonstrated in
\cite{Lazarian_etal:2015}, where the results of earlier simulations, e.g.
\cite{Beresnyak:2013}, were analyzed.

The present paper is a step forward in understanding the process of self-induced
fast reconnection.  Here we explore the nature of turbulence driving in the
reconnection region.  If tearing mode is absolutely essential for driving
turbulence, one may still argue that the actual reconnection is happening via
tearing, while the turbulence is playing only an auxiliary role for the process.
 Our results, in fact, testify that the actual picture is very different.  The
process of tearing mode plays in 3D a role at the earliest stage of
reconnection.  As the system evolves in time the outflows induced by the
reconnection region become turbulent, with Kelvin-Helmholtz instability playing
the dominant role.  As the reconnection grows, the region becomes more and more
turbulent with the tearing instability being overtaken or even suppressed, not
playing a role on the reconnection process overall.

Our simulations are performed in the high beta plasma regime and in such
conditions the reconnection outflow does not induce sufficient turbulence to
trigger the self-accelerating process of "reconnection instability"
\cite[see][]{LazarianVishniac:2009} though.

While the MHD simulations show a very different picture for 2D and 3D
self-driven reconnection, the particles-in-cell (PIC) simulations tend to show
similar tearing patterns both in 3D and 2D.  One possible explanation is related
to limitations of present-day PIC simulation, given that these do not present
enough particles in the reconnection regions to result in developed turbulence.
Therefore, in such "viscous" regime the Kelvin-Helmholtz instability is
suppressed and cannot operate and the only signatures that can be seen arise
from the tearing instability.  In other words, the "viscous" outflow does not
feel the additional degrees of freedom that would allow high Reynolds turbulent
behavior to take place.  Nevertheless, the high resolution PIC simulations
presented by Hui Li in a reconnection review by \cite{Lazarian_etal:2019} show
the signatures of developing turbulence, e.g., the Richardson dispersion of
magnetic field lines was reported.  Therefore, we expect that the results we now
obtained with MHD modelling can be also obtained/confirmed with very high
particle number PIC simulations.

There is, however, another puzzle that is presented by the comparison of the 3D
kinetic and MHD simulations.  The kinetic simulations show higher reconnection
rates and it is important to understand whether these differences persist for
reconnection at all scales or they are just a transient feature of reconnection
processes taking place for small-scale reconnection.  This issue was recently
addressed in \cite{Beresnyak:2018} using Hall-MHD code.  The results there
testify that the reconnection rates for self-driven 3D turbulent reconnection
obtained with Hall-MHD gradually converge to the results obtained for the 3D MHD
self-driven reconnection.  This is what one expects from the theory \cite[see
LV99;][]{Eyink_etal:2011, Eyink:2015}.  Nevertheless, in terms of our present
study the convergence of the results obtained with MHD and Hall-MHD code testify
that the results in the present paper will not change in the presence of
additional plasma effects.

Our confirmation of the predictions of turbulent reconnection theory formulated
in the LV99, and subsequent theoretical studies, also has bearing on the ongoing
discussion of the so called "reconnection-mediated turbulence" idea presented in
a number of theoretical papers \cite[see][]{LoureiroBoldyrev:2017a,
LoureiroBoldyrev:2017b, BoldyrevLoureiro:2017, BoldyrevLoureiro:2018,
Walker_etal:2018, Mallet_etal:2017a, Mallet_etal:2017b, Comisso_etal:2018,
Vech_etal:2018a}.  For sufficiently large Reynolds numbers, due to both the
process of "dynamical alignment" and the effect of magnetic fluctuations getting
more anisotropic with the decrease of the scale, the current sheets prone to
tearing instability can develop.  Such changes of the turbulence at the scale
$\lambda_c$ in the vicinity of the dissipation scale do not change the nature of
turbulent cascade, which lies on scales of the inertial range $\gg \lambda_c$.
Our study is therefore suggestive that if reconnection takes place at small
scales, comparable to $\lambda_c$ , it will also be turbulent as demonstrated by
our simulations.  However, since reconnection does not happen at the larger
eddies scales it is preferred to refer this hypothetical regime as
"tearing-mediated turbulence" instead.  The objective reality is, however, that
in the reconnection community historically only bursts of reconnection was
considered.

\section{Conclusions}
\label{sec:conclusions}

In this work we analyzed two MHD instabilities: tearing mode and
Kelvin--Helmholtz, which are candidates for processes responsible for turbulence
generation in spontaneous reconnection, e.g. the reconnection without externally
imposed turbulent driving.  The generated turbulence is due to the initially
imposed weak noise present in the vicinity of the Harris current sheet.  We
analyzed factors important for growth of both instabilities, but also those
which suppress them.  The analysis presented in this work has shown important
results which can be synthesized  in the following:

\begin{itemize}
\item The region of current sheet with the presence of initial noise develops
into a region with conditions favorable for development of MHD instabilities,
such as tearing mode or Kelvin--Helmholtz instability.  Although the tearing
instability is natural for thin elongated current sheets, the conditions for
Kelvin--Helmholtz instability have never been verified before in systems with
stochastic reconnection.

\item Evolution of stochastic reconnection leads to formation of shear regions,
both magnetic and velocity, with broad range of thicknesses and longitudinal
dimensions.  Our studies indicate that the regions of magnetic shear for later
times are somewhat thinner comparing to those of velocity shear, while the
longitudinal scales of these regions are shorter for velocity shear comparing to
the magnetic one.

\item Tearing instability is expected to develop at earlier stages, while, once
a sufficient amplitude of turbulence is generated near the current sheet, it can
be suppressed due to the presence of the transverse component of magnetic field
$B_n$.  As shown in \cite{SomovVerneta:1993}, for $\xi = B_n / B > S^{-3/4}$
this instability is suppressed.  We demonstrate that in our models $\xi$ can be
sufficiently large but does not completely suppress the instability.  Taking
into account the contribution of transverse component $B_n$, the instability can
develop with dynamical time shorter than the Alfvénic time $t_A$ under favorable
circumstances.  The estimated maximum growth rate $\gamma_{max}$, initially
below unity, reaches values of a few tens at later times with the most of
detected points laying in the range $0.1 - 10$.

\item Due to misalignment of the outflows from neighboring reconnection events,
they can generate enough sheared flows to induce Kelvin--Helmholtz instability.
The Mach numbers calculated with respect to the shear velocity $U$ satisfy
necessary conditions for the instability to develop.  Our analysis indicates the
presence of sheared regions with broad range of amplitudes, $10^{-2} \le U \le
1$, and thicknesses, $10^{-3} \le \delta \le 0.5$. The distributions of the
estimated maximum growth rates $\gamma_{max}$ peak at values between $1 - 10$
and reaching maximum values close to $100$, suggest the growth of
Kelvin--Helmholtz instability (within the dynamical time) much shorter than the
time $t_A$.
\end{itemize}

\acknowledgments
G.K. acknowledges support from CNPq (no. 304891/2016-9). D.F.G. thanks the
Brazilian agencies CNPq (no. 311128/2017-3) and FAPESP (no.  2013/10559-5) for
financial support.  A.L. acknowledges the NSF grant 1816234 and NASA TCAN
144AAG1967.  E.T.V. acknowledges the support of the AAS.  This work has made use
of the computing facilities the Academic Supercomputing Center in Krak\'ow,
Poland (Supercomputer Prometheus at ACK CYFRONET AGH) and of the Laboratory of
Astrophysics (EACH/USP, Brazil).

\bibliography{ms}

\begin{thebibliography}{}
\expandafter\ifx\csname natexlab\endcsname\relax\def\natexlab#1{#1}\fi
\providecommand{\url}[1]{\href{#1}{#1}}
\providecommand{\dodoi}[1]{doi:~\href{http://doi.org/#1}{\nolinkurl{#1}}}
\providecommand{\doeprint}[1]{\href{http://ascl.net/#1}{\nolinkurl{http://ascl.net/#1}}}
\providecommand{\doarXiv}[1]{\href{https://arxiv.org/abs/#1}{\nolinkurl{https://arxiv.org/abs/#1}}}

\bibitem[{{Alfv{\'e}n}(1942)}]{Alfven:1942}
{Alfv{\'e}n}, H. 1942, \nat, 150, 405, \dodoi{10.1038/150405d0}

\bibitem[{{Ara} {et~al.}(1978){Ara}, {Basu}, {Coppi}, {Laval}, {Rosenbluth}, \&
  {Waddell}}]{Ara_etal:1978}
{Ara}, G., {Basu}, B., {Coppi}, B., {et~al.} 1978, Annals of Physics, 112, 443,
  \dodoi{10.1016/S0003-4916(78)80007-4}

\bibitem[{{Armstrong} {et~al.}(1995){Armstrong}, {Rickett}, \&
  {Spangler}}]{Armstrong_etal:1995}
{Armstrong}, J.~W., {Rickett}, B.~J., \& {Spangler}, S.~R. 1995, \apj, 443,
  209, \dodoi{10.1086/175515}

\bibitem[{{Beresnyak}(2013)}]{Beresnyak:2013}
{Beresnyak}, A. 2013, ArXiv e-prints.
\newblock \doarXiv{1301.7424}

\bibitem[{{Beresnyak}(2017)}]{Beresnyak:2017}
---. 2017, \apj, 834, 47, \dodoi{10.3847/1538-4357/834/1/47}

\bibitem[{{Beresnyak}(2018)}]{Beresnyak:2018}
{Beresnyak}, A. 2018, in Journal of Physics Conference Series, Vol. 1031,
  Journal of Physics Conference Series, 012001

\bibitem[{{Berlok} \& {Pfrommer}(2019)}]{BerlokPfrommer:2019}
{Berlok}, T., \& {Pfrommer}, C. 2019, \mnras, 485, 908,
  \dodoi{10.1093/mnras/stz379}

\bibitem[{{Bhattacharjee} {et~al.}(2009){Bhattacharjee}, {Huang}, {Yang}, \&
  {Rogers}}]{Bhattacharjee_etal:2009}
{Bhattacharjee}, A., {Huang}, Y.-M., {Yang}, H., \& {Rogers}, B. 2009, Physics
  of Plasmas, 16, 112102, \dodoi{10.1063/1.3264103}

\bibitem[{{Boldyrev} \& {Loureiro}(2017)}]{BoldyrevLoureiro:2017}
{Boldyrev}, S., \& {Loureiro}, N.~F. 2017, \apj, 844, 125,
  \dodoi{10.3847/1538-4357/aa7d02}

\bibitem[{{Boldyrev} \& {Loureiro}(2018)}]{BoldyrevLoureiro:2018}
{Boldyrev}, S., \& {Loureiro}, N.~F. 2018, in Journal of Physics Conference
  Series, Vol. 1100, Journal of Physics Conference Series, 012003

\bibitem[{{Chandrasekhar}(1961)}]{Chandrasekhar:1961}
{Chandrasekhar}, S. 1961, {Hydrodynamic and hydromagnetic stability} ({Oxford})

\bibitem[{{Chen} {et~al.}(1997){Chen}, {Otto}, \& {Lee}}]{Chen_etal:1997}
{Chen}, Q., {Otto}, A., \& {Lee}, L.~C. 1997, \jgr, 102, 151,
  \dodoi{10.1029/96JA03144}

\bibitem[{{Chepurnov} {et~al.}(2015){Chepurnov}, {Burkhart}, {Lazarian}, \&
  {Stanimirovic}}]{Chepurnov_etal:2015}
{Chepurnov}, A., {Burkhart}, B., {Lazarian}, A., \& {Stanimirovic}, S. 2015,
  \apj, 810, 33, \dodoi{10.1088/0004-637X/810/1/33}

\bibitem[{{Chepurnov} \& {Lazarian}(2010)}]{ChepurnovLazarian:2010}
{Chepurnov}, A., \& {Lazarian}, A. 2010, \apj, 710, 853,
  \dodoi{10.1088/0004-637X/710/1/853}

\bibitem[{{Ciaravella} \& {Raymond}(2008)}]{CiaravellaRaymond:2008}
{Ciaravella}, A., \& {Raymond}, J.~C. 2008, \apj, 686, 1372,
  \dodoi{10.1086/590655}

\bibitem[{{Comisso} {et~al.}(2018){Comisso}, {Huang}, {Lingam}, {Hirvijoki}, \&
  {Bhattacharjee}}]{Comisso_etal:2018}
{Comisso}, L., {Huang}, Y.~M., {Lingam}, M., {Hirvijoki}, E., \&
  {Bhattacharjee}, A. 2018, \apj, 854, 103, \dodoi{10.3847/1538-4357/aaac83}

\bibitem[{{Coppi} {et~al.}(1976){Coppi}, {Galvao}, {Pellat}, {Rosenbluth}, \&
  {Rutherford}}]{Coppi_etal:1976}
{Coppi}, B., {Galvao}, R., {Pellat}, R., {Rosenbluth}, M., \& {Rutherford}, P.
  1976, Soviet Journal of Plasma Physics, 2, 533

\bibitem[{{Dougherty} {et~al.}(1989){Dougherty}, {Edelman}, \&
  {Hyman}}]{Dougherty_etal:1989}
{Dougherty}, R.~L., {Edelman}, A.~S., \& {Hyman}, J.~M. 1989, Mathematics of
  Computation, 52, 471–494, \dodoi{10.1090/S0025-5718-1989-0962209-1}

\bibitem[{{Drazin} \& {Reid}(1981)}]{DrazinReid:1981}
{Drazin}, P.~G., \& {Reid}, W.~H. 1981, NASA STI/Recon Technical Report A, 82,
  17950

\bibitem[{{Eyink} {et~al.}(2013){Eyink}, {Vishniac}, {Lalescu}, {Aluie},
  {Kanov}, {B{\"u}rger}, {Burns}, {Meneveau}, \& {Szalay}}]{Eyink_etal:2013}
{Eyink}, G., {Vishniac}, E., {Lalescu}, C., {et~al.} 2013, \nat, 497, 466,
  \dodoi{10.1038/nature12128}

\bibitem[{{Eyink}(2011)}]{Eyink:2011}
{Eyink}, G.~L. 2011, \pre, 83, 056405, \dodoi{10.1103/PhysRevE.83.056405}

\bibitem[{{Eyink}(2015)}]{Eyink:2015}
---. 2015, \apj, 807, 137, \dodoi{10.1088/0004-637X/807/2/137}

\bibitem[{{Eyink} {et~al.}(2011){Eyink}, {Lazarian}, \&
  {Vishniac}}]{Eyink_etal:2011}
{Eyink}, G.~L., {Lazarian}, A., \& {Vishniac}, E.~T. 2011, \apj, 743, 51,
  \dodoi{10.1088/0004-637X/743/1/51}

\bibitem[{{Faganello} \& {Califano}(2017)}]{FaganelloCalifano:2017}
{Faganello}, M., \& {Califano}, F. 2017, Journal of Plasma Physics, 83,
  535830601, \dodoi{10.1017/S0022377817000770}

\bibitem[{{Fejer}(1964)}]{Fejer:1964}
{Fejer}, J.~A. 1964, Physics of Fluids, 7, 499, \dodoi{10.1063/1.1711229}

\bibitem[{{Forbes}(2001)}]{Forbes:2001}
{Forbes}, T.~G. 2001, Earth, Planets, and Space, 53, 423,
  \dodoi{10.1186/BF03353252}

\bibitem[{{Frank} {et~al.}(1996){Frank}, {Jones}, {Ryu}, \&
  {Gaalaas}}]{Frank_etal:1996}
{Frank}, A., {Jones}, T.~W., {Ryu}, D., \& {Gaalaas}, J.~B. 1996, \apj, 460,
  777, \dodoi{10.1086/177009}

\bibitem[{{Furth} {et~al.}(1963){Furth}, {Killeen}, \&
  {Rosenbluth}}]{Furth_etal:1963}
{Furth}, H.~P., {Killeen}, J., \& {Rosenbluth}, M.~N. 1963, Physics of Fluids,
  6, 459, \dodoi{10.1063/1.1706761}

\bibitem[{{Galeev} \& {Zeleny{\v{i}}}(1975)}]{GaleevZelenyi:1975}
{Galeev}, A.~A., \& {Zeleny{\v{i}}}, L.~M. 1975, Soviet Journal of Experimental
  and Theoretical Physics Letters, 22, 170

\bibitem[{{Galeev} \& {Zeleny{\v{i}}}(1976)}]{GaleevZelenyi:1976}
---. 1976, Soviet Journal of Experimental and Theoretical Physics, 43, 1113

\bibitem[{{Gerwin}(1968)}]{Gerwin:1968}
{Gerwin}, R.~A. 1968, Reviews of Modern Physics, 40, 652,
  \dodoi{10.1103/RevModPhys.40.652}

\bibitem[{{Goldreich} \& {Sridhar}(1995)}]{GoldreichSridhar:1995}
{Goldreich}, P., \& {Sridhar}, S. 1995, \apj, 438, 763, \dodoi{10.1086/175121}

\bibitem[{{Gonz{\'a}lez-Casanova} {et~al.}(2018){Gonz{\'a}lez-Casanova},
  {Lazarian}, \& {Cho}}]{Gonzales-Casanova_etal:2018}
{Gonz{\'a}lez-Casanova}, D.~F., {Lazarian}, A., \& {Cho}, J. 2018, \mnras, 475,
  3324, \dodoi{10.1093/mnras/sty006}

\bibitem[{{Greco} {et~al.}(2008){Greco}, {Chuychai}, {Matthaeus}, {Servidio},
  \& {Dmitruk}}]{Greco_etal:2008}
{Greco}, A., {Chuychai}, P., {Matthaeus}, W.~H., {Servidio}, S., \& {Dmitruk},
  P. 2008, \grl, 35, L19111, \dodoi{10.1029/2008GL035454}

\bibitem[{{Gudkov} \& {Troshichev}(1996)}]{GudkovTroshichev:1996}
{Gudkov}, M.~G., \& {Troshichev}, O.~A. 1996, Journal of Atmospheric and
  Terrestrial Physics, 58, 613, \dodoi{10.1016/0021-9169(95)00082-8}

\bibitem[{{Huang} \& {Bhattacharjee}(2016)}]{HuangBhattacharjee:2016}
{Huang}, Y.-M., \& {Bhattacharjee}, A. 2016, \apj, 818, 20,
  \dodoi{10.3847/0004-637X/818/1/20}

\bibitem[{{Jacobson} \& {Moses}(1984)}]{JacobsonMoses:1984}
{Jacobson}, A.~R., \& {Moses}, R.~W. 1984, \pra, 29, 3335,
  \dodoi{10.1103/PhysRevA.29.3335}

\bibitem[{{Khabarova} \& {Obridko}(2012)}]{KhabarovaObridko:2012}
{Khabarova}, O., \& {Obridko}, V. 2012, \apj, 761, 82,
  \dodoi{10.1088/0004-637X/761/2/82}

\bibitem[{Kiefer(1953)}]{Kiefer:1953}
Kiefer, J. 1953, Proceedings of the American Mathematical Society, 4, 502,
  \dodoi{10.1090/S0002-9939-1953-0055639-3}

\bibitem[{{Kowal} {et~al.}(2017){Kowal}, {Falceta-Gon{\c c}alves}, {Lazarian},
  \& {Vishniac}}]{Kowal_etal:2017}
{Kowal}, G., {Falceta-Gon{\c c}alves}, D.~A., {Lazarian}, A., \& {Vishniac},
  E.~T. 2017, \apj, 838, 91, \dodoi{10.3847/1538-4357/aa6001}

\bibitem[{{Kowal} {et~al.}(2009){Kowal}, {Lazarian}, {Vishniac}, \&
  {Otmianowska-Mazur}}]{Kowal_etal:2009}
{Kowal}, G., {Lazarian}, A., {Vishniac}, E.~T., \& {Otmianowska-Mazur}, K.
  2009, \apj, 700, 63, \dodoi{10.1088/0004-637X/700/1/63}

\bibitem[{{Kowal} {et~al.}(2012){Kowal}, {Lazarian}, {Vishniac}, \&
  {Otmianowska-Mazur}}]{Kowal_etal:2012}
---. 2012, Nonlinear Processes in Geophysics, 19, 297,
  \dodoi{10.5194/npg-19-297-2012}

\bibitem[{{Lazarian}(2005)}]{Lazarian:2005}
{Lazarian}, A. 2005, in American Institute of Physics Conference Series, Vol.
  784, Magnetic Fields in the Universe: From Laboratory and Stars to Primordial
  Structures., ed. E.~M. {de Gouveia dal Pino}, G.~{Lugones}, \& A.~{Lazarian},
  42--53

\bibitem[{{Lazarian} {et~al.}(2012){Lazarian}, {Esquivel}, \&
  {Crutcher}}]{Lazarian_etal:2012}
{Lazarian}, A., {Esquivel}, A., \& {Crutcher}, R. 2012, \apj, 757, 154,
  \dodoi{10.1088/0004-637X/757/2/154}

\bibitem[{{Lazarian} {et~al.}(2019){Lazarian}, {Eyink}, {Jafari}, {Kowal},
  {Li}, {X}, \& {Vishniac}}]{Lazarian_etal:2019}
{Lazarian}, A., {Eyink}, G., {Jafari}, A., {et~al.} 2019, Physics of Plasmas

\bibitem[{{Lazarian} {et~al.}(2015){Lazarian}, {Eyink}, {Vishniac}, \&
  {Kowal}}]{Lazarian_etal:2015}
{Lazarian}, A., {Eyink}, G., {Vishniac}, E., \& {Kowal}, G. 2015, Philosophical
  Transactions of the Royal Society of London Series A, 373, 20140144,
  \dodoi{10.1098/rsta.2014.0144}

\bibitem[{{Lazarian} \& {Vishniac}(1999)}]{LazarianVishniac:1999}
{Lazarian}, A., \& {Vishniac}, E.~T. 1999, \apj, 517, 700,
  \dodoi{10.1086/307233}

\bibitem[{{Lazarian} \& {Vishniac}(2009)}]{LazarianVishniac:2009}
{Lazarian}, A., \& {Vishniac}, E.~T. 2009, in Revista Mexicana de Astronomia y
  Astrofisica Conference Series, Vol.~36, Magnetic Fields in the Universe II:
  From Laboratory and Stars to the Primordial Universe, ed. A.~{Esquivel},
  J.~{Franco}, G.~{García-Segura}, E.~M. {de Gouveia Dal Pino}, A.~{Lazarian},
  S.~{Lizano}, \& A.~Raga, 81--88

\bibitem[{{Le{\~a}o} {et~al.}(2013){Le{\~a}o}, {de Gouveia Dal Pino},
  {Santos-Lima}, \& {Lazarian}}]{Leao_etal:2013}
{Le{\~a}o}, M.~R.~M., {de Gouveia Dal Pino}, E.~M., {Santos-Lima}, R., \&
  {Lazarian}, A. 2013, \apj, 777, 46, \dodoi{10.1088/0004-637X/777/1/46}

\bibitem[{{Loureiro} \&
  {Boldyrev}(2017{\natexlab{a}})}]{LoureiroBoldyrev:2017a}
{Loureiro}, N.~F., \& {Boldyrev}, S. 2017{\natexlab{a}}, \prl, 118, 245101,
  \dodoi{10.1103/PhysRevLett.118.245101}

\bibitem[{{Loureiro} \&
  {Boldyrev}(2017{\natexlab{b}})}]{LoureiroBoldyrev:2017b}
---. 2017{\natexlab{b}}, \apj, 850, 182, \dodoi{10.3847/1538-4357/aa9754}

\bibitem[{{Loureiro} {et~al.}(2007){Loureiro}, {Schekochihin}, \&
  {Cowley}}]{Loureiro_etal:2007}
{Loureiro}, N.~F., {Schekochihin}, A.~A., \& {Cowley}, S.~C. 2007, Physics of
  Plasmas, 14, 100703, \dodoi{10.1063/1.2783986}

\bibitem[{{Loureiro} {et~al.}(2013){Loureiro}, {Schekochihin}, \&
  {Uzdensky}}]{Loureiro_etal:2013}
{Loureiro}, N.~F., {Schekochihin}, A.~A., \& {Uzdensky}, D.~A. 2013, \pre, 87,
  013102, \dodoi{10.1103/PhysRevE.87.013102}

\bibitem[{{Mallet} {et~al.}(2017{\natexlab{a}}){Mallet}, {Schekochihin}, \&
  {Chand ran}}]{Mallet_etal:2017b}
{Mallet}, A., {Schekochihin}, A.~A., \& {Chand ran}, B. D.~G.
  2017{\natexlab{a}}, Journal of Plasma Physics, 83, 905830609,
  \dodoi{10.1017/S0022377817000812}

\bibitem[{{Mallet} {et~al.}(2017{\natexlab{b}}){Mallet}, {Schekochihin}, \&
  {Chandran}}]{Mallet_etal:2017a}
{Mallet}, A., {Schekochihin}, A.~A., \& {Chandran}, B.~D.~G.
  2017{\natexlab{b}}, \mnras, 468, 4862, \dodoi{10.1093/mnras/stx670}

\bibitem[{{Matthaeus} \& {Lamkin}(1985)}]{MatthaeusLamkin:1985}
{Matthaeus}, W.~H., \& {Lamkin}, S.~L. 1985, Physics of Fluids, 28, 303,
  \dodoi{10.1063/1.865147}

\bibitem[{{Matthaeus} \& {Lamkin}(1986)}]{MatthaeusLamkin:1986}
---. 1986, Physics of Fluids, 29, 2513, \dodoi{10.1063/1.866004}

\bibitem[{{Michalke}(1964)}]{Michalke:1964}
{Michalke}, A. 1964, Journal of Fluid Mechanics, 19, 543,
  \dodoi{10.1017/S0022112064000908}

\bibitem[{{Miura} \& {Pritchett}(1982)}]{MiuraPritchett:1982}
{Miura}, A., \& {Pritchett}, P.~L. 1982, \jgr, 87, 7431,
  \dodoi{10.1029/JA087iA09p07431}

\bibitem[{{Oishi} {et~al.}(2015){Oishi}, {Mac Low}, {Collins}, \&
  {Tamura}}]{Oishi_etal:2015}
{Oishi}, J.~S., {Mac Low}, M.-M., {Collins}, D.~C., \& {Tamura}, M. 2015,
  \apjl, 806, L12, \dodoi{10.1088/2041-8205/806/1/L12}

\bibitem[{{Ong} \& {Roderick}(1972)}]{OngRoderick:1972}
{Ong}, R.~S.~B., \& {Roderick}, N. 1972, \planss, 20, 1,
  \dodoi{10.1016/0032-0633(72)90135-3}

\bibitem[{{Padoan} {et~al.}(2009){Padoan}, {Juvela}, {Kritsuk}, \&
  {Norman}}]{Padoan_etal:2009}
{Padoan}, P., {Juvela}, M., {Kritsuk}, A., \& {Norman}, M.~L. 2009, \apjl, 707,
  L153, \dodoi{10.1088/0004-637X/707/2/L153}

\bibitem[{{Parker}(1957)}]{Parker:1957}
{Parker}, E.~N. 1957, \jgr, 62, 509, \dodoi{10.1029/JZ062i004p00509}

\bibitem[{{Petschek}(1964)}]{Petschek:1964}
{Petschek}, H.~E. 1964, NASA Special Publication, 50, 425

\bibitem[{Rayleigh(1879)}]{Rayleigh:1879}
Rayleigh, L. 1879, Proceedings of the London Mathematical Society, s1-11, 57,
  \dodoi{10.1112/plms/s1-11.1.57}

\bibitem[{{Santos-Lima} {et~al.}(2013){Santos-Lima}, {de Gouveia Dal Pino}, \&
  {Lazarian}}]{Santos-Lima_etal:2013}
{Santos-Lima}, R., {de Gouveia Dal Pino}, E.~M., \& {Lazarian}, A. 2013,
  \mnras, 429, 3371, \dodoi{10.1093/mnras/sts597}

\bibitem[{{Santos-Lima} {et~al.}(2010){Santos-Lima}, {Lazarian}, {de Gouveia
  Dal Pino}, \& {Cho}}]{Santos-Lima_etal:2010}
{Santos-Lima}, R., {Lazarian}, A., {de Gouveia Dal Pino}, E.~M., \& {Cho}, J.
  2010, \apj, 714, 442, \dodoi{10.1088/0004-637X/714/1/442}

\bibitem[{{Schindler}(1974)}]{Schindler:1974}
{Schindler}, K. 1974, \jgr, 79, 2803, \dodoi{10.1029/JA079i019p02803}

\bibitem[{{Sen}(1964)}]{Sen:1964}
{Sen}, A.~K. 1964, Physics of Fluids, 7, 1293, \dodoi{10.1063/1.1711374}

\bibitem[{{Servidio} {et~al.}(2011){Servidio}, {Dmitruk}, {Greco}, {Wan},
  {Donato}, {Cassak}, {Shay}, {Carbone}, \& {Matthaeus}}]{Servidio_etal:2011}
{Servidio}, S., {Dmitruk}, P., {Greco}, A., {et~al.} 2011, Nonlinear Processes
  in Geophysics, 18, 675, \dodoi{10.5194/npg-18-675-2011}

\bibitem[{{Somov} \& {Verneta}(1989)}]{SomovVerneta:1989}
{Somov}, B.~V., \& {Verneta}, A.~I. 1989, \solphys, 120, 93,
  \dodoi{10.1007/BF00148536}

\bibitem[{{Somov} \& {Verneta}(1993)}]{SomovVerneta:1993}
---. 1993, \ssr, 65, 253, \dodoi{10.1007/BF00754510}

\bibitem[{{Sweet}(1958)}]{Sweet:1958}
{Sweet}, P.~A. 1958, The Observatory, 78, 30

\bibitem[{{Sych} {et~al.}(2015){Sych}, {Karlick{\'y}}, {Altyntsev},
  {Dud{\'\i}k}, \& {Kashapova}}]{Sych_etal:2015}
{Sych}, R., {Karlick{\'y}}, M., {Altyntsev}, A., {Dud{\'\i}k}, J., \&
  {Kashapova}, L. 2015, \aap, 577, A43, \dodoi{10.1051/0004-6361/201424834}

\bibitem[{{Sych} {et~al.}(2009){Sych}, {Nakariakov}, {Karlicky}, \&
  {Anfinogentov}}]{Sych_etal:2009}
{Sych}, R., {Nakariakov}, V.~M., {Karlicky}, M., \& {Anfinogentov}, S. 2009,
  \aap, 505, 791, \dodoi{10.1051/0004-6361/200912132}

\bibitem[{{Syrovatskii}(1981)}]{Syrovatskii:1981}
{Syrovatskii}, S.~I. 1981, \araa, 19, 163,
  \dodoi{10.1146/annurev.aa.19.090181.001115}

\bibitem[{{Takamoto}(2018)}]{Takamoto:2018}
{Takamoto}, M. 2018, \mnras, 476, 4263, \dodoi{10.1093/mnras/sty493}

\bibitem[{{Takamoto} {et~al.}(2015){Takamoto}, {Inoue}, \&
  {Lazarian}}]{Takamoto_etal:2015}
{Takamoto}, M., {Inoue}, T., \& {Lazarian}, A. 2015, \apj, 815, 16,
  \dodoi{10.1088/0004-637X/815/1/16}

\bibitem[{{Vech} {et~al.}(2018){Vech}, {Mallet}, {Klein}, \&
  {Kasper}}]{Vech_etal:2018a}
{Vech}, D., {Mallet}, A., {Klein}, K.~G., \& {Kasper}, J.~C. 2018, \apjl, 855,
  L27, \dodoi{10.3847/2041-8213/aab351}

\bibitem[{{Walker}(1981)}]{Walker:1981}
{Walker}, A.~D.~M. 1981, \planss, 29, 1119,
  \dodoi{10.1016/0032-0633(81)90011-8}

\bibitem[{{Walker} {et~al.}(2018){Walker}, {Boldyrev}, \&
  {Loureiro}}]{Walker_etal:2018}
{Walker}, J., {Boldyrev}, S., \& {Loureiro}, N.~F. 2018, \pre, 98, 033209,
  \dodoi{10.1103/PhysRevE.98.033209}

\end{thebibliography}

\end{document}